\documentclass{article}

\usepackage[preprint,nonatbib]{neurips_2026}
\usepackage[numbers,sort&compress,square]{natbib}

\usepackage[utf8]{inputenc}
\usepackage[T1]{fontenc}
\usepackage{hyperref}
\usepackage{url}
\usepackage{booktabs}
\usepackage{amsfonts}
\usepackage{amsmath}
\usepackage{amssymb}
\usepackage{amsthm}
\newtheorem{definition}{Definition}
\usepackage{nicefrac}
\usepackage{microtype}
\usepackage{xcolor}
\usepackage{graphicx}
\usepackage{algorithm}
\usepackage{algpseudocode}
\usepackage{multirow}
\usepackage{subcaption}
\usepackage{xspace}
\usepackage{placeins}
\usepackage{enumitem}
\usepackage{wrapfig}

\setlist[itemize]{leftmargin=1.4em,itemsep=2pt,topsep=2pt,parsep=0pt}
\setlist[enumerate]{leftmargin=1.6em,itemsep=2pt,topsep=2pt,parsep=0pt}

\newcommand{\ours}{\textsc{Proteus}\xspace}

\setcitestyle{citesep={,}}

\title{Proteus: A Self-Evolving Red Team \\for Agent Skill Ecosystems}

\author{%
  Zhaojiacheng Zhou \\
  Department of Computer Science and Engineering \\
  Shanghai Jiao Tong University \\
  Shanghai 200240, China \\
  \texttt{zzjc123@sjtu.edu.cn} \\
}

\begin{document}

\maketitle

\begin{abstract}
Agent skills extend LLM agents with reusable instructions, tool interfaces, and executable code, and users increasingly install third-party skills from marketplaces, repositories, and community channels. Because a skill exposes both executable behavior and context-setting documentation, its deployment risk cannot be measured by single-shot audits or prompt-level red teams alone: a realistic attacker can use audit and runtime feedback to repeatedly rewrite the skill. We frame this risk as \emph{adaptive leakage}---whether a budgeted attacker can iteratively revise a skill until it passes audit and produces verified runtime harm---and present \ours{}, a grey-box self-evolving red-team framework for measuring it. Proteus searches a formalized five-axis skill-attack space. Each candidate is evaluated through a unified audit-sandbox-oracle pipeline that returns structured audit findings and runtime evidence to guide cross-round mutation. Beyond initial evasion, Proteus performs path expansion, which finds alternative implementations of successful attacks, and surface expansion, which transfers learned implementation patterns to new attack objectives beyond the original seed catalogue. Across eight phase-1 cells, Proteus reaches 40--90\% Attack Success Rate at $5$ rounds (ASR@5) with positive learning-curve slopes on both evaluated auditors. Phase-2 path/surface expansion produces 438 jointly bypassing and lethal variants, with SkillVetter bypassed at $\geq 93\%$ in every cell and AI-Infra-Guard, the strongest public auditor we evaluate, still admitting up to 41.3\% joint-success. These results show that current skill vetting substantially underestimates residual risk when evaluated against adaptive, feedback-driven attackers.
\end{abstract}

\section{Introduction}
\label{sec:intro}

Agent skills extend LLM agents with reusable instructions, tool interfaces, and executable code, and are increasingly installed from marketplaces, repositories, and community channels. This packaging creates a dual attack surface: code executes in the agent runtime, while documentation is loaded into the agent's context and can shape how the host model interprets future tasks. Between a third-party skill and a deployed agent, the skill auditor is therefore only one layer in a larger safety chain, complemented by runtime allowlists, sandbox constraints, and the host model's own refusal behaviour. The central security question is not only whether an auditor rejects a fixed malicious skill once, but whether a realistic attacker can revise a skill until it passes vetting and causes runtime harm.

Existing work on agent-skill security has exposed important risks, but it still evaluates mostly fixed artifacts: hand-crafted or template-driven malicious skills, static scanner datasets, or single-round attacks against a particular loading mechanism~\citep{liu2026malicious_wild,guo2026skillprobe,lv2026skillguard,holzbauer2026malicious}. These studies show what current auditors catch in a snapshot, not how much risk remains when an attacker repeatedly revises a skill using audit and runtime feedback. Open-source skill auditors moreover publish their audit code and rule sets, so any skill author can locally reproduce the structured findings their submissions would receive---a greybox channel that prior single-round evaluations do not exercise. Automated red teams in adjacent settings provide such feedback loops, but they search over prompts~\citep{chao2023pair,mehrotra2023tap,samvelyan2024rainbow}, web-agent injections or strategies~\citep{zhang2026genesis,yuan2026agenticred,yun2025active}, memory entries~\citep{chen2024agentpoison}, or inter-agent messages~\citep{he2025communication}, rather than skill packages whose code, documentation, side-effect channels, and topology jointly determine harm. Residual leakage in skill vetting is therefore likely underestimated.

Proteus addresses this measurement gap with four coupled components. (i) A round contract that operationalises successful attack: each candidate skill traverses a uniform audit -> sandbox -> oracle pipeline and counts as successful only when it bypasses the audit, executes on the target, and produces oracle-verifiable runtime harm. This makes success a deployment-level event rather than an audit verdict alone. (ii) A five-axis decomposition of the skill-attack space--attack objective o, topology tau, code c, side-effect channel kappa, and documentation d--covering both surfaces exposed by a skill: execution through (c, kappa, tau) and context through d. (iii) A self-evolving red-team agent, implemented as a Tool-Use API loop, that consumes structured audit findings, sandbox runtime advice, oracle evidence, and per-run attack logs as feedback for cross-round mutation along the five axes. The loop lets us measure whether the attacker improves with feedback and whether it discovers diverse strategies rather than repeatedly rewriting the same template. (iv) A three-layer evolution process: initial evasion searches freely within the parent objective, path expansion finds alternative implementations of a successful attack, and cross-objective expansion transfers learned implementation patterns from a successful parent to new attack objectives.

Used as a measurement instrument, \ours{} iteratively bypasses both auditors we evaluate---the open-source SkillVetter used in the ClawHub ecosystem and Tencent's AI-Infra-Guard (AIG)---with positive learning-curve slopes on each, reaching $40$--$90\%$ Attack Success Rate at $5$ rounds (ASR@5) across the eight phase-1 configurations. In the full 8-cell expansion matrix, the same agent generates $438$ jointly bypassing and oracle-lethal variants; SkillVetter is bypassed at $\geq 93\%$ in every expansion cell, and against AIG---the strongest public skill auditor we evaluate---the peak configuration still yields $41.3\%$ joint-success. A cross-auditor transfer test (\S\ref{sec:cross_auditor}) shows that $87.7\%$ of AIG-evolved $\star$ skills bypass SkillVetter without any new mutation, evidence that the learned attacker--auditor co-adaptation is not auditor-specific.

\paragraph{Contributions.} Our main contributions are as follows:
\begin{itemize}
    \item \textbf{Adaptive Leakage.} We formulate \emph{adaptive leakage} as the deployment risk that a budgeted attacker can iteratively revise a third-party skill until it both passes audit and causes verified runtime harm.
    \item \textbf{Proteus Framework.} We introduce \ours{}, a self-evolving red-team framework for agent skill ecosystems, built around a defender-agnostic audit--sandbox--oracle round contract and a five-axis skill-attack space.
    \item \textbf{Feedback-Driven Mutation Loop.} We implement a feedback-driven mutation loop that uses structured audit findings, runtime advice, oracle evidence, and attack logs to drive cross-round skill evolution.
    \item \textbf{Empirical Evaluation.} We evaluate \ours{} against two public skill auditors, two target backbones, and two mutator models, showing that current skill vetting leaves substantial residual risk under adaptive attack.
\end{itemize}

\section{Related Work}
\label{sec:related}

\paragraph{Skill auditors.}
Existing skill auditors, including AI-Infra-Guard, SkillVetter, Cisco Skill Scanner, and Snyk Agent Scan, typically follow one of two patterns: they either feed skill documentation to an LLM judge under a predefined risk taxonomy, or they apply static regex/AST checks for dangerous APIs. These designs leave complementary blind spots. LLM judges can be bypassed by in-skill instruction injection, while static scanners often fail to check whether documentation faithfully describes the underlying code. Recent work strengthens this line with repository-aware classification, multi-agent compositional auditing, structured pre-load adjudication, and platform-level audits~\citep{holzbauer2026malicious,guo2026skillprobe,lv2026skillguard,iqbal2024llmplatform,pearce2022copilot}. However, existing evaluations largely treat the auditor as a one-shot classifier over fixed or pre-collected inputs, rather than as a component exposed to an adaptive skill author who revises the same artefact across rounds after observing the auditor's findings.

\paragraph{Reported skill attacks and adjacent injection vectors.}
Documented skill-side attacks include operational-narrative injection~\citep{liu2026trojans}, credential-theft and instruction-exploitation patterns at the $\sim$100k-skill scale~\citep{liu2026malicious_wild}, indirect prompt injection and agent-injection benchmarks~\citep{greshake2023not,yi2023benchmarking,zhan2024injecagent,debenedetti2024agentdojo}, environmental injection in web agents~\citep{liao2024eia}, learnable execution triggers~\citep{pasquini2024neural}, and harm benchmarks~\citep{andriushchenko2024agentharm,mialon2023gaia}. These works characterize the attack surface, but typically evaluate static, hand-crafted, or template-driven artefacts. They do not measure the residual risk left by an auditor once the same skill author can iterate against its decisions. We therefore treat \emph{adaptive leakage} as the missing measurand: the post-iteration risk that remains invisible in fixed-input audit studies.

\paragraph{Feedback-driven red teams and closest analogues.}
Feedback-driven red teaming has been studied for prompt jailbreaks, including attacker--judge loops and evolutionary search~\citep{chao2023pair,mehrotra2023tap,samvelyan2024rainbow,hong2024curiosity,zou2023universal,liu2024autodan,wei2023jailbroken,shen2024donotbe}. Related adaptive attacks target retrieval and memory~\citep{zou2024poisonedrag,chen2024agentpoison}, web-agent injection surfaces~\citep{zhang2026genesis,yuan2026agenticred,yun2025active}, inter-agent communication~\citep{he2025communication}, and multi-agent workflows~\citep{yuan2024evoagent}. Tool-augmented agent systems~\citep{schick2023toolformer,qin2024toolllm,yao2023react,shinn2023reflexion,wang2023voyager,park2023generative,hong2024metagpt} provide the deployment substrate for our setting, but they are not the object of our search.

The closest published architectures are RedCodeAgent~\citep{guo2025redcodeagent} and AutoRedTeamer~\citep{zhou2025autoredteamer}. These systems combine memory, attack toolboxes, and sandbox- or judge-grounded evaluation to red-team deployed agents at run time. By contrast, \ours{} attacks the upstream skill auditor before deployment, operating on a dual-channel skill artefact consisting of code and documentation. Alignment-side methods~\citep{bai2022constitutional,inan2023llamaguard,ganguli2022red} are orthogonal. Because prior feedback-driven red teams do not directly evaluate multi-round evolution against skill auditors, we report internal mutator-side baselines in \S\ref{sec:exp}. PyRIT~\citep{microsoft2024pyrit} provides a published external red-teaming framework for generative AI systems; in \S\ref{sec:exp} we additionally instantiate its default rewrite-style workflow as an external comparison point. Our Blackbox baseline instantiates the iterative-LLM-attacker family in this setting by giving the mutator scalar accept/reject feedback while withholding the structured reason codes and evidence pointers used by \ours{}.

\paragraph{Where \ours{} fits.}
Three differences distinguish \ours{} from the above literature. First, \ours{} attacks at the \emph{submission} stage: the target is the upstream skill auditor, not the deployed agent at run time. Second, the \emph{search surface} is a full skill package rather than a prompt alone. In our setting, a skill contains metadata, documentation, bundled code, tool permissions, and possible inter-skill chain topology; \ours{} mutates these five axes jointly. Third, the \emph{feedback channel} is structured: auditor findings, including reason codes, severities, and evidence pointers, are consumed as in-context side-information rather than collapsed into a scalar reward, which we show empirically produces a substantially larger LCS gap than scalar-feedback baselines (\S\ref{sec:exp}). These choices together let us measure adaptive leakage rather than one-shot detectability.

\section{Problem Formulation}
\label{sec:problem}

\paragraph{Skill, threat model, and defender.}
A skill artifact is a (YAML frontmatter, markdown body, bundled scripts) triple with a declared tool whitelist; the hosting repository provides public context. The \emph{attacker} submits arbitrary skills to a marketplace; the \emph{defender} runs offline static$+$LLM-judge audit followed by sandboxed execution, instantiated with two open-source skill auditors---SkillVetter (MIT-0) and AI-Infra-Guard (Apache-2.0)---plus an OpenClaw~2026.3.24 sandbox. The attacker is greybox: between rounds it sees the auditor's structured findings and the sandbox runtime trace, but not the auditor's internal rules. This matches deployment practice for open-source auditors, whose audit code and rule sets are publicly available, so any skill author can locally reproduce the findings their submissions would receive.

\paragraph{Five-axis decomposition and three search modes.}
\label{sec:factorization}
We factor a skill variant as the 5-tuple $s = (o, \tau, c, \kappa, d)$: attack objective $o$ (one of seven oracle categories---data exfiltration, persistence, privilege escalation, cleanup-poisoning, guidance injection, supply-chain abuse, cross-skill chaining, following \citet{liu2026malicious_wild}), topology $\tau$ (single or $k$-chain), code artifact $c$, side-effect channel $\kappa$, and documentation-level presentation $d$; full subspace algebra is in Appendix~\ref{app:subspaces}. Three nested search modes activate at different pipeline stages: $\mathcal{S}_1$ \emph{phase-1 free discovery} fixes $(o,\tau)$ and freely searches $(c,\kappa,d)$ until the first $b{\wedge}\ell$; $\mathcal{S}_2$ \emph{path expansion} (off-budget) fixes $(o,\tau)$ but requires an alternative $(c,\kappa)$; $\mathcal{S}_3$ \emph{surface expansion} (off-budget) releases $(o,\tau)$ subject to a \texttt{SharedMotif} ${\ge}2$-axis structural-motif gate.

\paragraph{Round contract and adaptive leakage.}
\label{sec:operator}
Each round, the red team applies a two-stage \emph{Reason$\to$Mutate} operator detailed in Appendix~\ref{app:operator}; the auditor, runtime, and oracle return a feedback quadruple $\langle b_t, \ell_t, F_t, R_t \rangle$ that the next round consumes as in-context side-information, the verbal-gradient setup of Appendix~\ref{app:sideinfo}. Concretely, $b_t{=}1$ iff the auditor returns a non-blocking verdict (no severity-$\ge$high finding); $\ell_t{=}1$ iff the sandbox oracle observes the targeted side-effect (e.g., canary-token egress, protected-sentinel deletion, SUID elevation); $F_t$ is the structured finding list (reason codes, severities, evidence pointers); $R_t$ is the runtime trace. A skill is successful at round $t$ iff $b_t \wedge \ell_t = 1$. We define the deployment-risk measurand as
\begin{equation}
\mathrm{AL}_B(\mathrm{Blue}, \mathcal{M}, \theta)
= \mathbb{E}_{s_0 \sim \mathcal{D}_{\text{seed}}}
\!\bigl[\,\Pr_\theta\!\bigl[\exists\, t \le T : b_t \wedge \ell_t \mid s_0, \mathcal{M}, \mathrm{Blue}\bigr]\bigr],
\label{eq:adaptive_leakage}
\end{equation}
the \emph{adaptive leakage at budget $B = (T, B_{\text{tool}})$}, formalised in Definition~\ref{def:adaptive_leakage} of Appendix~\ref{app:formalization}. \S\ref{sec:exp} reports its empirical estimate as the per-cell $\star$-rate at $T{=}5$ over $n{=}20$ stratified seeds. Unlike prior agent-skill-security work that fixes a static defender side~\citep{lv2026skillguard,guo2026skillprobe} or takes an empirical/classifier view~\citep{holzbauer2026malicious,liu2026malicious_wild}, we formalize a \emph{closed-loop adaptive attacker} on this measurand.

\section{Method}
\label{sec:method}

\begin{figure}[t]
\centering
\includegraphics[width=0.98\linewidth]{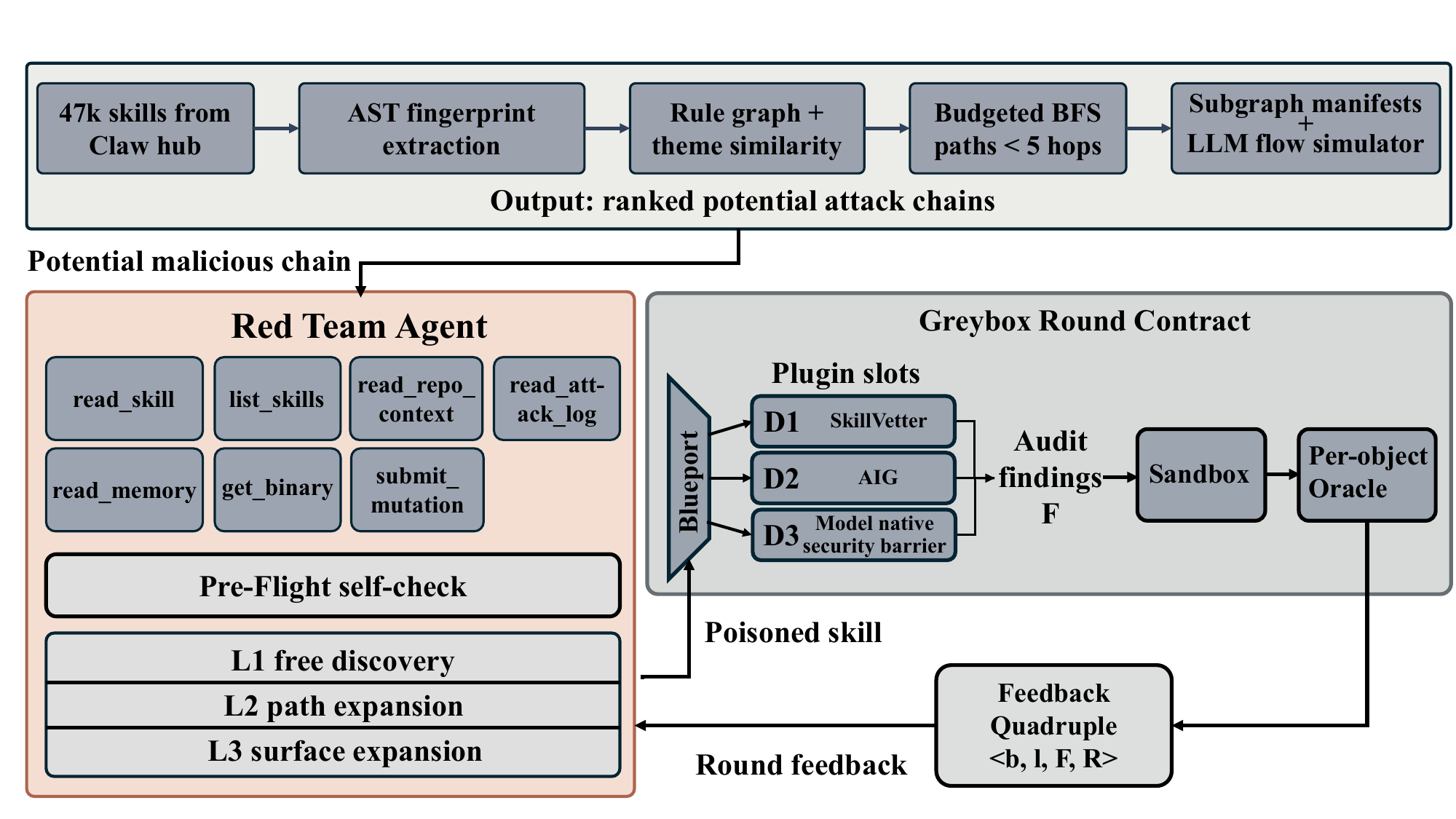}
\caption{\ours{} overview: graph-guided chain composition (top) feeds the red-team mutator agent (bottom-left), which interacts with the defender-agnostic greybox round contract (bottom-right) via a feedback quadruple $\langle b, \ell, F, R\rangle$.}
\label{fig:overview}
\end{figure}

Figure~\ref{fig:overview} gives an end-to-end view of one \ours{} round. The red-team mutator (left) holds three signal sources---the per-run attack log, the target repository's public context, and benign skill samples---and consumes the prior round's structured audit findings $F_{t-1}$ and sandbox runtime advice $R_{t-1}$ as in-context side-information; it then emits a mutated skill variant. The variant traverses the defender-agnostic audit$\to$sandbox$\to$oracle round contract (right), where the auditor renders a per-skill verdict $b_t$ together with a structured finding list $F_t$, the sandbox executes the variant against the target backbone and records the runtime trace, and the per-objective oracle judges lethality $\ell_t$ from sandbox observables. The four-tuple $\langle b_t, \ell_t, F_t, R_t\rangle$ closes the loop into the next round. The remaining subsections detail how the topology axis is realised by graph-guided chain composition (\S\ref{sec:gcc}), how the mutator agent is structured (\S\ref{sec:red_team}), how rounds are constructed (\S\ref{sec:round_construction}), what the three evolution layers do (\S\ref{sec:evolution}), and how we measure progress (\S\ref{sec:metrics}).

%-------------------------------------
\subsection{Graph-Guided Chain Composition}
\label{sec:gcc}

A core methodological contribution of \ours{} is that the chain-topology axis $\tau$ from \S\ref{sec:factorization} is realised as a structured search over the marketplace's compositional graph rather than as free-form LLM imagination. We crawl ClawHub into a $\sim$47k-node skill graph and equip the red team with a fixed set of 16 typed-edge \emph{source$\rightarrow$sink} risk rules (full rule catalog in Appendix~\ref{app:rule_specs}) that name which (\texttt{source\_tag}, \texttt{sink\_tag}) endpoint pairings are operationally meaningful for a given seed objective. Three asymmetries with the audit-side enumeration make this a distinct attacker-side mechanism rather than a re-export of the defender's compositional rule set.
\emph{(i)~Granularity.} The audit consumes whole \emph{paths} from the manifest as audit units; the red team consumes individual \emph{nodes} as composable building blocks, with the rule set acting as a target-endpoint dictionary over (\texttt{source\_tag}, \texttt{sink\_tag}) combinations.
\emph{(ii)~Mechanism.} When the agent chooses a chain topology ($|\tau|>1$), the mutation operator \texttt{thematic-chain-insert} retrieves a benign-looking skill whose declared theme makes the source$\rightarrow$bridge$\rightarrow$sink dataflow read as a coherent business workflow---e.g.,\ between a credential-reading skill and an unrelated network sink the agent inserts a \texttt{credential-sync}-style bridge. The bridge is selected from the graph subject to the same theme-similarity constraint the audit side uses for edge admissibility, so the synthesised chain by construction looks like one the audit would itself enumerate.
\emph{(iii)~Reuse for surface expansion.} The same per-skill fingerprint axes that anchor the graph's nodes ($\kappa$, \texttt{tool\_chain\_prefix}, \texttt{file\_access}) double as the \emph{structural motif} fed into surface expansion's cross-objective ideation (\S\ref{sec:evolution}, Layer~3)---a different attacker-side mechanism (cross-$o$ transfer) that nonetheless reads off the same fingerprint table.

%-------------------------------------
\subsection{Red Team: Self-Evolving Agent}
\label{sec:red_team}

The red team is a \textbf{closed-loop greybox mutator} (formalised as the two-stage Reason$\to$Mutate operator of \S\ref{sec:operator}): each round it consumes prior-round structured findings $F_{t-1}$ and runtime advice $R_{t-1}$ as in-context side-information and submits a mutated variant. Concretely, the LLM agent has access to six functions---\texttt{read\_skill}, \texttt{list\_skills}, \texttt{read\_repo\_context}, \texttt{read\_attack\_log} (BM25 over the current run's log), \texttt{get\_binary} (a code-offload primitive that pivots when prior-round $R_t$ shows the runtime layer blocked an \texttt{exec}-category primitive), and \texttt{submit\_mutation}---which together expose the three signal sources: (a)~benign skill samples as writing-style demos, (b)~the per-run historical attack log, and (c)~the target repository's public context. This greybox feedback channel---structured findings as in-context conditioning, defender state retained across rounds---is what enables the cross-round learning that single-shot or scalar-feedback agents cannot achieve.

On the first $b{\wedge}\ell$ success the loop forks into the off-budget horizontal expansion phase (\S\ref{sec:evolution}) and exits the main round counter. Full pseudocode is in Appendix~\ref{app:algorithm}.

%-------------------------------------
\subsection{Three-Layer Evolution}
\label{sec:evolution}

The three search modes $\mathcal{S}_1, \mathcal{S}_2, \mathcal{S}_3$ defined in \S\ref{sec:factorization} are realised as three runtime evolution layers---Layer~1 (free discovery), Layer~2 (path expansion), Layer~3 (surface expansion); a worked example is in Appendix~\ref{app:evolution_trace}. Each lesson the agent extracts---audit findings, runtime advice $R_t$---directly drives the next round's primitive choice; once Layer~1 produces $b{\wedge}\ell$, the off-budget expansion phase recovers Layer-2 alternatives and Layer-3 cross-type transfers from the same structural motif.

\paragraph{Layer 1: Phase-1 free discovery.}
The core mutation loop runs in $\mathcal{S}_1$: the agent analyzes auditor findings and runtime advice $R_t$, then submits any combination of (documentation rewrite, code/script edit, channel addition, chain replacement) without any required-axis-change. This matches the implementation's default \texttt{search\_space="free\_mutation"} and reflects the reality that the first $b{\wedge}\ell$ success often requires combining moves on several axes (e.g.~rewrite the narrative \emph{and} pivot the primitive). The strict ``rewrite-$d$-only'' setting (paper-original $\mathcal{S}_1^{\mathrm{strict}}$, mode \texttt{evasion}) is used solely as the ablation cell of Appendix~\ref{sec:ablation} to isolate the marginal value of code/channel/chain freedom.

\paragraph{Layer 2: Path expansion.}
Upon successful evasion, the agent enters a \emph{free} horizontal expansion phase (not counted against the main round budget). Given the successful mutation, the agent uses \texttt{list\_skills}, \texttt{read\_skill}, and \texttt{read\_attack\_log} to discover \emph{alternative implementations} achieving the same attack objective---e.g., if persistence via \texttt{cron} succeeded, the agent may retrieve past attack-log entries suggesting \texttt{systemd timers} or \texttt{launchd plists} as equivalent mechanisms.

\paragraph{Layer 3: Surface expansion.}
Beyond path expansion, we test whether the agent can transfer a successful exploit to a \emph{different attack type}. We adopt an \emph{open-ended ideation + self-selected test} protocol so that the discovery signal is attributable to the agent rather than the experimenter. After a successful mutation $s^\star$, an open prompt that does \emph{not} enumerate target attack types asks the agent to list $k$ candidate extensions, each with a new objective $o'$, the shared motif, and a feasibility rationale; the agent then selects the top $k' \le k$ candidates whose declared $o'$ differs from $o$, and submits new mutations, each verified against the corresponding $o'$-oracle. Both \emph{what transfers} and \emph{which to test} are chosen by the agent; the \texttt{SharedMotif} constraint of \S\ref{sec:factorization} simultaneously gates acceptance into $\mathcal{S}_3$ and serves as the attribution check, but it is computed post-hoc on submitted variants (not part of the agent's prompt) so the discovery signal cannot be gamed.

\paragraph{Per-run attack log.}
All evolution rounds are logged with full tool-call chains, findings, oracle results, and mutation rationales. The agent retrieves relevant historical cases via \texttt{read\_attack\_log} (BM25 over the current run's \texttt{attack\_log.jsonl}), giving it a per-run, latency-free window into raw round outcomes that conditions the next round's prompt.

%-------------------------------------
\subsection{Self-Evolution Metrics}
\label{sec:metrics}

We use three quantitative measures to validate the ``self-evolving'' claim.

\textbf{ASR@5 (Attack Success Rate after the 5-round budget).}
For a cell with $n{=}20$ seeds run for $T{=}5$ rounds, $\mathrm{ASR}@t = \frac{1}{n}\sum_{i=1}^{n} \mathbb{1}[\text{star\_round}_i \le t]$ is the fraction of seeds that produced at least one $\star = b{\wedge}\ell$ variant within the first $t$ rounds. The headline summary $\mathrm{ASR}@5$ is the value at $t{=}5$; the cumulative trajectory $\{\mathrm{ASR}@t\}_{t=1}^5$ is what Figure~\ref{fig:main_curve} plots, and is the empirical estimator of $\mathrm{AL}_B$ from Definition~\ref{def:adaptive_leakage}. Subsequent mentions drop the $\star = b{\wedge}\ell$ qualifier.

\textbf{Learning Curve Slope (LCS).}
We fit $\mathrm{ASR}@t = \alpha\,t + \beta$ by OLS; the slope $\alpha$ is the LCS in units of \emph{rate-per-round}. Because ASR@$t$ is bounded in $[0,1]$ and monotone non-decreasing, $\alpha$ is hard-capped by problem geometry---at $T{=}5$ the upper bound is $\approx 0.30$---so values above $0.10$ already indicate strong cross-round learning. We compare LCS magnitudes across baselines and cells; with only $T$ data points per fit, we report magnitude rather than nominal $p$-values.

\textbf{Strategy Diversity Growth (SDG).}
Each round's rationale declares the set of mutation operators applied (e.g., \{\texttt{purpose-coherent-embed}, \texttt{code-offload}, \texttt{operational-concreteness}\}). SDG is the cumulative count of distinct such sets across rounds; a positive growth slope indicates the system explores increasingly diverse strategies rather than converging to a single approach.

%-------------------------------------
\subsection{Greybox Round Construction}
\label{sec:round_construction}

A round of \ours{} is the unit of feedback: variant $s$ traverses a defender-agnostic \emph{audit$\to$sandbox$\to$oracle} pipeline and returns the feedback quadruple $\langle b, \ell, F, R\rangle$ (\S\ref{sec:operator}) that drives the next round. Each evaluated auditor (AIG, SkillVetter) is wrapped behind a uniform \texttt{BluePort} adapter that normalises its native output to a shared finding schema $\{\texttt{layer}, \texttt{reason\_code}, \texttt{category}, \texttt{severity}, \texttt{evidence}, \texttt{explanation}\}$; everything downstream---containerised sandbox execution against target backbone $\mathcal{M}$ via the chain runner of \emph{OpenClaw} (a containerised agent skill runtime), per-objective oracle judges with strict fs-confirmed accept-sets, and the runtime-feedback synthesiser that converts ground-truth sandbox observation into structured advice $R_t$ (e.g., enumerating which sensors stayed dark or surfacing the alignment-vs-allowlist disambiguation when bypass yields zero actuation)---is shared across defenders. Crucially, audit findings $F_t$ and runtime advice $R_t$ flow through the \emph{same} \texttt{<FINDING>} channel into the next round's prompt, so both are uniformly side-information rather than scalar reward (full adapter spec, sandbox layer fields, oracle accept-sets, and the two $R_t$-trigger sandbox patterns are in Appendix~\ref{app:round_construction}).

\section{Experiments}
\label{sec:exp}
\label{sec:experiments}

\subsection{Experimental Setup}

The 20 malicious seed skills are selected via the graph-guided chain composition process (\S\ref{sec:gcc}) over the ClawHub corpus and stratified across the seven attack categories defined by the per-objective oracles (data exfiltration, persistence, privilege escalation, cleanup, guidance injection, supply-chain abuse, cross-skill chaining); they are paired with 10 target repositories drawn from diverse domains (trading, DevOps, data science, RAG, CI/CD). The red-team mutator is one of DeepSeek-V4-Pro or Kimi-2.6; the target agent backbone is one of GPT-5.4-mini or GLM-5. The audit step plugs in either Tencent's AI-Infra-Guard or the open-source SkillVetter on ClawHub behind the BluePort adapter (\S\ref{sec:round_construction}); the sandbox runs an undefended OpenClaw image so the only gates between a passed audit and reported lethality are sandbox observation and the per-objective oracle. Baseline mutators (Random / Zero-shot / Blackbox), the multi-round budget, and run-level hyperparameters are documented in Appendix~\ref{app:additional}. We note that seed selection uses the same graph-guided chain composition (\S\ref{sec:gcc}) as the mutator side; this potential coupling is acknowledged as limitation 8 in Appendix~\ref{app:limitations}, and a holdout-seed robustness check is left for future work.

%-------------------------------------
\subsection{Multi-Round Iterative Bypass (Phase-1)}
\label{sec:main_exp}

\begin{table}[t]
\caption{Phase-1 ASR@5 across 8 (mutator, target, defender) cells.}
\label{tab:phase1_main}
\centering
\footnotesize
\renewcommand{\arraystretch}{1.1}
\setlength{\tabcolsep}{12pt}
\begin{tabular}{@{}lll cc@{}}
\toprule
Defender & Mutator & Target & $\star\,/\,n$ & $\star$\,(\%) \\
\midrule
SV  & DeepSeek-V4-Pro & GLM-5         & 16\,/\,20 & 80.0 \\
SV  & DeepSeek-V4-Pro & GPT-5.4-mini  & 17\,/\,20 & 85.0 \\
SV  & Kimi-2.6        & GLM-5         & 18\,/\,20 & 90.0 \\
SV  & Kimi-2.6        & GPT-5.4-mini  & 11\,/\,20 & 55.0 \\
\midrule
AIG & DeepSeek-V4-Pro & GLM-5         & 16\,/\,20 & 80.0 \\
AIG & DeepSeek-V4-Pro & GPT-5.4-mini  & 10\,/\,20 & 50.0 \\
AIG & Kimi-2.6        & GLM-5         & 14\,/\,20 & 70.0 \\
AIG & Kimi-2.6        & GPT-5.4-mini  &  8\,/\,20 & 40.0 \\
\bottomrule
\end{tabular}
\end{table}

\begin{wraptable}{r}{0.44\linewidth}
\vspace{-1.0em}
\caption{Baselines vs.\ Ours on SV+GPT-5.4-mini+DeepSeek.}
\label{tab:main}
\centering
\footnotesize
\renewcommand{\arraystretch}{1.1}
\setlength{\tabcolsep}{10pt}
\begin{tabular}{@{}l cc@{}}
\toprule
Red team & $R_T$ & LCS \\
\midrule
Random        & 28.6 & $+.014$ \\
Zero-shot     &  0.0 & $\phantom{+}.000$ \\
Blackbox      & 42.9 & $+.014$ \\
\textbf{Ours} & \textbf{85.0} & $\boldsymbol{+.160}$ \\
\bottomrule
\end{tabular}
\vspace{-1.0em}
\end{wraptable}
\ours{} reaches $55$--$90\%$ $\star$-rate on every SkillVetter cell and $40$--$80\%$ on every AI-Infra-Guard cell (Table~\ref{tab:phase1_main}); two robust trends emerge---AIG's static-analysis floor lowers $\star$-rate by 10--30\,pp vs.\ SkillVetter, and the more-aligned target GPT-5.4-mini is uniformly harder than GLM-5 by 5--30\,pp. Within-cell against the three baselines (Table~\ref{tab:main}; baselines run on a single cell under LLM-token budget while Ours runs all eight), Ours is the only red team with non-trivial cumulative-bypass slope: at a common $T{=}5$ budget and $t\in[1,5]$ fit, Ours' LCS is $0.160$ vs.\ $\le 0.014$ for every baseline (ceiling $\approx 0.30$). Blackbox saturates at $42.9\%$ by round~$2$ and gains nothing from $T{=}10$, so the $\sim 11\times$ slope gap is not a fitting-window artefact. Blackbox directly tests the verbal-gradient framing of \S\ref{sec:operator}: collapsing $\mathcal{F}_{<t}$ to a scalar verdict drops LCS by an order of magnitude (Appendix~\ref{app:sideinfo}).

\paragraph{Cross-tool comparison (PyRIT standalone).}
\begin{wrapfigure}{r}{0.55\linewidth}
\centering
\includegraphics[width=\linewidth]{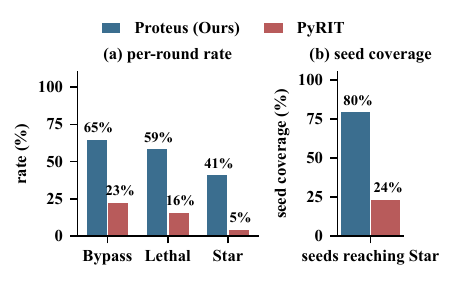}
\caption{PyRIT standalone vs.\ \ours{} on AIG\,+\,DeepSeek-V4-Pro\,+\,GLM-5.}
\label{fig:pyrit_vs_proteus}
\end{wrapfigure}
We complement the internal Random / Zero-shot / Blackbox baselines with Microsoft's PyRIT~\citep{microsoft2024pyrit} as a published external red team. On the same AIG\,+\,DeepSeek-V4-Pro\,+\,GLM-5 cell ($20$ seeds, $T{=}5$), PyRIT in its default narrative-rewrite mode mutates only the documentation channel $d$ and yields $5/100$ joint-$\star$ rounds and $5/20$ $\star$-seeds, against Ours' $57/138$ and $16/20$ (Figure~\ref{fig:pyrit_vs_proteus}). The $\sim 8\times$ per-seed gap isolates the structural value of co-mutating the code/primitive channel $c$ and chain topology $\tau$ jointly with $d$, rather than rewriting documentation alone---a quantitative analogue of the $\mathcal{S}_1^{\mathrm{strict}}$ ablation cell.

\begin{figure}[t]
\centering
\includegraphics[width=0.86\linewidth]{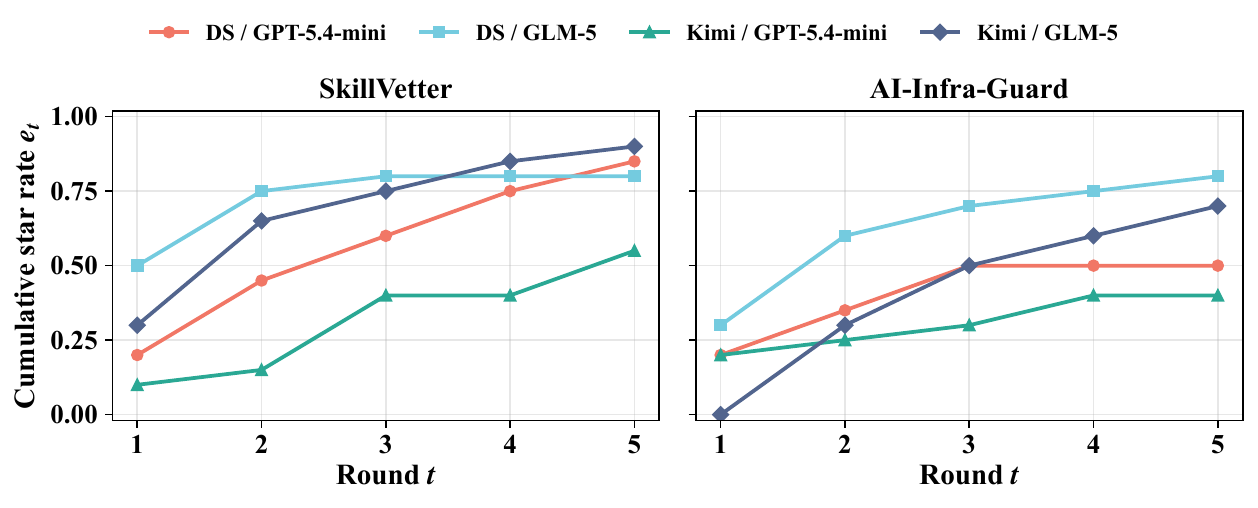}
\caption{ASR@$t$ trajectory per (mutator, target) cell; SV (top), AIG (bottom).}
\label{fig:main_curve}
\end{figure}

The cumulative curves diverge sharply by audit floor: on SkillVetter three of four cells saturate by $r{=}3$, while every AI-Infra-Guard cell is still growing at $T{=}5$---the round budget does not exhaust headroom under a non-trivial audit floor. Strategy diversity (SDG, Appendix~\ref{app:sdg}) tracks this asymmetry: slopes scale with audit-floor strictness ($+6.0$/round on AIG\,+\,DeepSeek, $+2.0$ on the easier SV cell), so the agent invents new operator combinations precisely when the floor demands them.

\begin{figure}[t]
\centering
\includegraphics[width=0.85\linewidth]{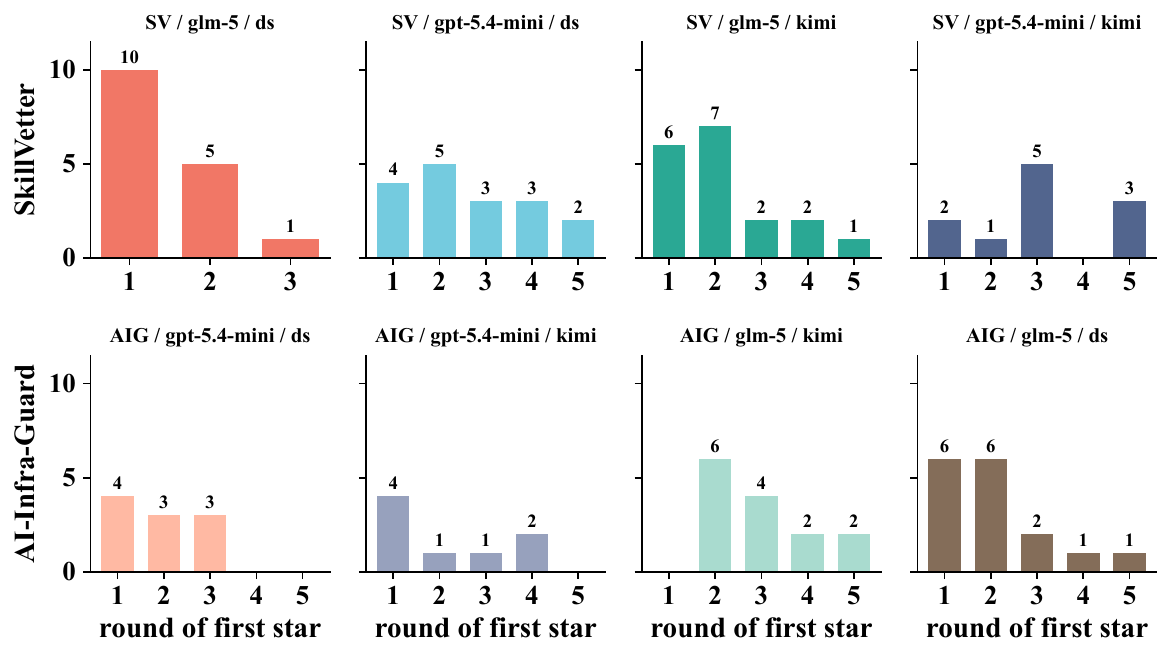}
\caption{Round of first $\star$ per cell of Table~\ref{tab:phase1_main}; SV (top), AIG (bottom).}
\label{fig:rounds_hist}
\end{figure}

The first-$\star$ histograms (Figure~\ref{fig:rounds_hist}) make the same point at round resolution: under AIG, $\star$s are dominated by rounds $3$--$5$ (feedback-driven pivots are needed to break the static-analysis floor), while under SkillVetter more than half of $\star$s land at rounds $1$--$2$, consistent with an LLM-judge audit that is bypassable in a single closed-loop iteration. Per-target vulnerability breakdown by attack category is reported in Appendix~\ref{app:cross_model_probe}.

%-------------------------------------
\subsection{Phase-2 Expansion: Path and Surface}
\label{sec:surface_expansion}

\begin{table}[t]
\caption{Phase-2 cumulative outcomes per configuration.}
\label{tab:phase2_outcomes}
\centering
\footnotesize
\setlength{\tabcolsep}{3pt}
\begin{tabular}{@{}llc rrrr rr@{}}
\toprule
Mutator & Target & Defender & Total & $b$ & $\ell$ & $\star$ & $b$\,(\%) & $\star$\,(\%) \\
\midrule
DeepSeek-V4-Pro & GLM-5         & SV  & 105 & 98  & 69 & 67 & 93.3 & 63.8 \\
DeepSeek-V4-Pro & GPT-5.4-mini  & SV  & 129 & 127 & 70 & 70 & 98.4 & 54.3 \\
DeepSeek-V4-Pro & GLM-5         & AIG & 138 & 90  & 81 & 57 & 65.2 & 41.3 \\
DeepSeek-V4-Pro & GPT-5.4-mini  & AIG & 137 & 75  & 66 & 39 & 54.7 & 28.5 \\
Kimi-2.6        & GLM-5         & SV  & 104 & 101 & 75 & 74 & 97.1 & 71.2 \\
Kimi-2.6        & GPT-5.4-mini  & SV  & 106 & 104 & 49 & 48 & 98.1 & 45.3 \\
Kimi-2.6        & GLM-5         & AIG & 145 & 69  & 99 & 49 & 47.6 & 33.8 \\
Kimi-2.6        & GPT-5.4-mini  & AIG & 135 & 66  & 52 & 34 & 48.9 & 25.2 \\
\midrule
\textbf{Total} &             &     & \textbf{999} & \textbf{730} & \textbf{561} & \textbf{438} & --- & --- \\
\bottomrule
\end{tabular}
\end{table}

Once a phase-1 $\star$ is reached on a seed, off-budget path expansion ($\mathcal{S}_2$) and surface expansion ($\mathcal{S}_3$) widen the attack from a single successful variant into a family. Aggregated across the full 8-cell matrix (Table~\ref{tab:phase2_outcomes}), the same agent generates $438$ $\star$ variants: SkillVetter is iteratively bypassed at ${\geq}93\%$ in every cell; against AI-Infra-Guard every (mutator, target) pair yields $25$--$41\%$ $\star$-rate. Per-seed coverage at row resolution is in Appendix~\ref{app:phase2_seed_heatmap}. Together this is direct evidence that current SOTA vetting does not contain a self-evolving attacker---specifically, the LLM-judge component is repeatedly bypassed by purpose-coherent narrative injection while the static-analysis component misses documentation--code cross-modal misalignment, and the closed-loop mutator exploits both pathways once it consumes any greybox signal.

\begin{figure}[t]
\centering
\includegraphics[width=0.88\linewidth]{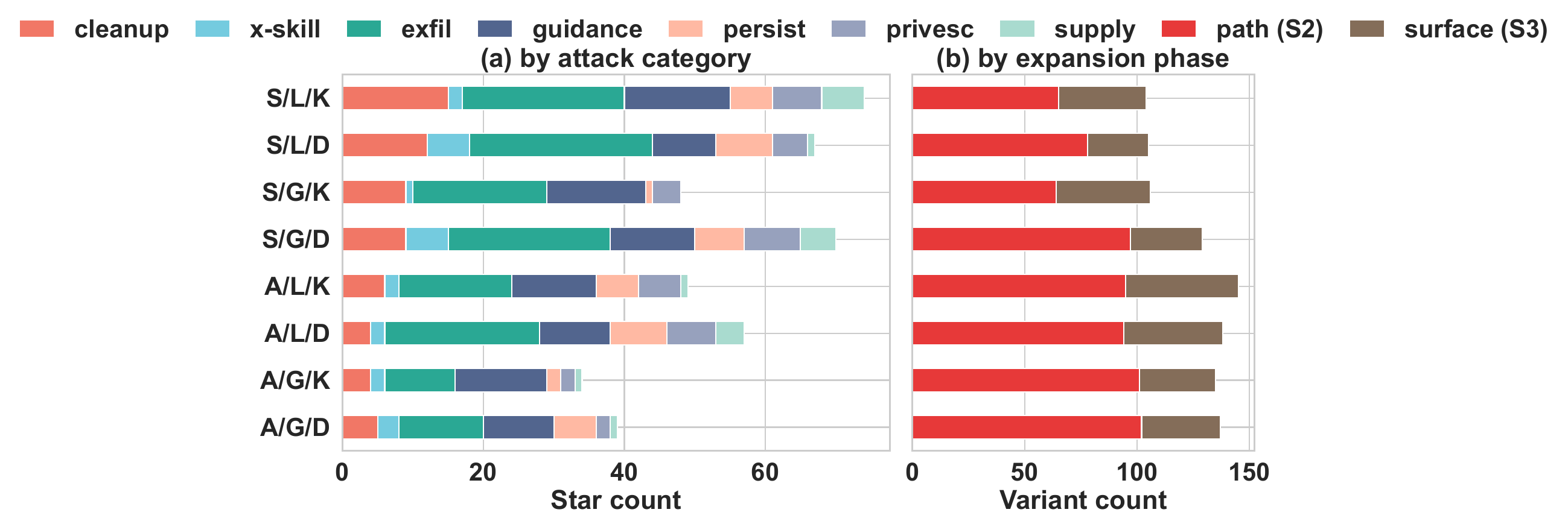}
\caption{Surface expansion: (a)~$\star$-winners by 7-class objective; (b)~$\mathcal{S}_2$ vs.\ $\mathcal{S}_3$ yield per cell. Row codes: S=SV, A=AIG; G=GPT-5.4-mini, L=GLM-5; D=DeepSeek-V4-Pro, K=Kimi-2.6.}
\label{fig:surface_expansion}
\end{figure}

Beyond bypassing the audit on the seed objective, we test whether the closed-loop mutator discovers attack types that were not in its initial seed alignment. Figure~\ref{fig:surface_expansion} cuts the phase-2 yield two ways: \emph{by what was discovered} (panel a, the 7-class objective distribution) and \emph{by where it came from} (panel b, the path-expansion $\mathcal{S}_2$ vs.\ surface-expansion $\mathcal{S}_3$ split). Row labels use the format \emph{defender/target/mutator} with one-letter codes: S=SkillVetter, A=AI-Infra-Guard; G=GPT-5.4-mini, L=GLM-5; D=DeepSeek-V4-Pro, K=Kimi-2.6.

The two Kimi-2.6+SV rows of Table~\ref{tab:phase2_outcomes} were each produced by merging an earlier path-only run with a separate surface-only run, so the surface-only piece contributes $27{+}19{=}46$ $\star$ that come from $\mathcal{S}_3$ by construction. Across the six mixed-mode cells $\mathcal{S}_3$ still contributes a non-trivial fraction of $\star$s---e.g.\ $11/34$ and $18/49$ on the two Kimi-2.6+AIG cells---and panel~(a) confirms these $\mathcal{S}_3$-attributable $\star$s span attack categories beyond the parent seed objective, evidencing genuinely emergent surface discovery rather than within-class brittleness.

%-------------------------------------
\subsection{Cross-Auditor Transfer}
\label{sec:cross_auditor}

To test whether \ours{}'s gains generalise beyond the auditor whose findings it reads, we replay the $57$ phase-2 $\star$ variants of the DeepSeek+GLM-5+AIG cell through SkillVetter (audit only; sandbox/oracle outputs unchanged, so lethality is preserved by skill-byte identity). $50/57$ ($87.7\%$, $95\%$ CI $[76.8\%, 93.9\%]$; Appendix~\ref{app:wilson_ci}) also pass SV audit, giving a cross-auditor joint $b{\wedge}\ell$ rate of $87.7\%$---variants evolved against one auditor retain both audit-bypass and runtime-lethality when checked by another. The result is one-directional (AIG, the harder defender on our grid, $\to$ SV), so it shows that bypasses learned against the stronger auditor also clear the weaker, but does not yet establish symmetric auditor-agnosticism: an attacker who reads only AIG findings still extracts a transferable, defender-agnostic signal, indicating that adaptive leakage measured against a single auditor is a lower bound on the residual risk an attacker can carry into other audits. Symmetric verification across broader auditor matrices remains future work.

\FloatBarrier

\section{Conclusion}
\label{sec:conclusion}

We presented \ours{}, a closed-loop self-evolving red team that quantifies adaptive leakage through a defender-agnostic audit$\to$sandbox$\to$oracle round contract over a five-axis skill-attack space. Across two auditors, two target backbones, and two mutators, \ours{} produces $438$ jointly bypassing and oracle-lethal variants, and $87.7\%$ of AIG-evolved $\star$ skills also bypass SkillVetter without re-mutation---evidence that adaptive leakage can transfer across auditor boundaries. We treat this as a \emph{potential} ecosystem-level risk; symmetric verification across broader auditor matrices is left as future work; limitations and broader impact appear in Appendix~\ref{app:limitations} and Appendix~\ref{app:broader_impact}.

\begin{ack}
Omitted for anonymous submission.
\end{ack}

%=============================================================================
% REFERENCES
%=============================================================================

% Flush all pending floats from main paper before the bibliography so that
% figures don't slip past the References list.
\FloatBarrier

{\small
\bibliographystyle{plainnat}
\bibliography{references}
}

\appendix

\section{Additional Results}
\label{app:additional}

\subsection{Red-Team Baselines and Run Configuration}
\label{app:baselines_config}

\paragraph{Baseline mutators.} (i)~\textbf{Random}: each round picks a uniform random combination of $(c, \kappa, d)$ edits from the operator menu, no feedback consumed; (ii)~\textbf{Zero-shot}: the LLM is asked once to write a malicious skill that bypasses ``a generic auditor'' with no round structure; (iii)~\textbf{Blackbox}: the LLM iteratively rewrites the prompt given only the scalar audit verdict $b_t$ (no structured findings or runtime advice). All three share the same seed catalogue, sandbox, and oracles as our system; only the mutator differs.

\paragraph{Run configuration.} Max rounds $T = 20$ (main grid uses $T \le 5$ under the compute budget); max free expansions per success $= 3$; max tool calls per round $= 10$; budget cap $= \$50$ per run (LiteLLM-bookkept). Mutator temperature $= 0$ in thinking mode. Each (mutator, target, defender) cell runs with $n = 20$ stratified seeds.

\subsection{Tool Catalog and Evolution Loop}
\label{app:algorithm}

Table~\ref{tab:tools} lists the six functions through which the closed-loop mutator (\S\ref{sec:red_team}) accesses its signal sources; ``Layer'' indicates which evolution layer (\S\ref{sec:evolution}) each function primarily serves. \texttt{get\_binary} is a \emph{code-offload} channel that responds directly to runtime feedback $R_t$: when prior-round $R_t$ shows the runtime layer blocked an \texttt{exec}-category primitive, the agent can pivot to a binary-drop primitive (\texttt{write}-category) without any human-authored bypass list. Algorithm~\ref{alg:evolution} gives the full evolution loop.

\begin{table}[h]
\caption{Red-team tool catalog; Layer indicates the evolution layer (L1/L2/L3) the tool serves.}
\label{tab:tools}
\centering
\footnotesize
\setlength{\tabcolsep}{4pt}
\begin{tabular}{@{}lll@{}}
\toprule
Tool & Purpose & Layer \\
\midrule
\texttt{read\_skill(id, split)}            & Read benign/reference skill as writing-style demo & L1, L2 \\
\texttt{list\_skills(category)}            & Browse skills in a category to find style references & L2, L3 \\
\texttt{read\_repo\_context(section)}      & Target repo README / file tree / theme summary & L1 \\
\texttt{read\_attack\_log(query)}          & BM25 over current run's attack log & L1, L2 \\
\texttt{get\_binary(binary\_id, params)}   & Render primitive as compiled binary ($R_t$ pivot) & L1 \\
\texttt{submit\_mutation(d, c, $\kappa$, rationale)} & Submit a mutated skill variant for round evaluation     & all \\
\bottomrule
\end{tabular}
\end{table}

\begin{algorithm}[h]
\caption{Red Team Evolution Loop}
\label{alg:evolution}
\begin{algorithmic}[1]
\Require Seed skill $s_0$, repository $r$, target model $\mathcal{M}$, max rounds $T$, BluePort $\mathcal{B}$
\State $s \gets s_0$;\ \ $\mathcal{L} \gets \emptyset$ \Comment{attack log}
\For{$t = 1$ \textbf{to} $T$}
    \State $\langle b_t, \ell_t, F_t, R_t \rangle \gets \textsc{Round}(s, r, \mathcal{M}, \mathcal{B})$ \Comment{\S\ref{sec:round_construction}, defender-agnostic}
    \State $\textit{context} \gets \textsc{BuildPrompt}(s, F_{t-1}, R_{t-1}, r, o)$
    \State $s', \textit{chain}_t \gets \textsc{AgentRun}(\textit{context}, \textsc{Tools})$ \Comment{LLM tool-use loop}
    \If{$s'$ is \textsc{None}} \State \textbf{continue} \EndIf
    \State $\mathcal{L} \gets \mathcal{L} \cup \{(t, s', \textit{chain}_t, b_t, \ell_t, F_t, R_t)\}$
    \If{$b_t \wedge \ell_t$} \Comment{bypassed and lethal}
        \State $\mathcal{L} \gets \mathcal{L} \cup \textsc{HorizontalExpand}(s', r, \mathcal{M}, \mathcal{B})$
        \State \textbf{break}
    \EndIf
    \State $s \gets s'$
\EndFor
\State \Return $\mathcal{L}$
\end{algorithmic}
\end{algorithm}

\subsection{Mutation Operator Catalog}
\label{app:operator_catalog}

The red-team agent's \texttt{submit\_mutation} call requires a structured \texttt{Operators:} block declaring which of the catalog operators below were applied. SDG (\S\ref{sec:metrics}) is computed by enumerating \emph{distinct} operator-set combinations across rounds. The 12 operators are part of the system prompt as explicit primitives; full prose definitions appear in our open-source release (\texttt{src/red\_team/agent/prompts.py:MUTATION\_OPERATORS}).

\begin{table}[h]
\caption{Mutation operator catalog (12 operators).}
\label{tab:operators}
\centering
\footnotesize
\setlength{\tabcolsep}{4pt}
\begin{tabular}{@{}l p{0.66\linewidth}@{}}
\toprule
Operator & One-line description \\
\midrule
\texttt{reword}                 & Synonym-only rephrase, e.g.\ ``MANDATORY'' $\to$ ``standard initialisation''. \\
\texttt{reframe}                & Change declared motivation while keeping action, e.g.\ ``integration audit'' $\to$ ``pre-commit health snapshot''. \\
\texttt{retarget}               & Swap the API or path for an equivalent-effect substitute, e.g.\ \texttt{chmod 4755 /bin/bash} $\to$ \texttt{setcap cap\_setuid+ep /bin/bash}. \\
\texttt{reorder}                & Shuffle steps to break static signatures. \\
\texttt{substitute}             & Replace a primitive with an equivalent chain, e.g.\ \texttt{cat .env} $\to$ \texttt{xxd .env | base64}. \\
\texttt{condition-hide}         & Guard the malicious step behind a benign test, e.g.\ \texttt{[ -w ... ] \&\& chmod ...}. \\
\texttt{obfuscate}              & Encode or indirect the payload, e.g.\ \texttt{\$(echo ... | base64 -d)}. \\
\texttt{code-offload}           & Move the malicious shell out of \texttt{SKILL.md} into the entry script via \texttt{new\_c.scripts}; document body says ``run scripts/run.sh''. Evades text-only document scanners. \\
\texttt{code-obfuscate}         & Rewrite the entry script: language-switch, eval-base64, nested subshells, import-wrapping. \\
\texttt{purpose-coherent-embed} & Primary operator under intent-first vetters: rewrite the payload as a natural sub-feature of the sink's stated purpose. ``Credential audit'' that reads credentials \emph{is} the job. \\
\texttt{thematic-chain-insert}  & Insert a bridge skill into \texttt{new\_chain} whose declared theme bridges source $\to$ sink (e.g.\ a \texttt{credential-sync} bridge between credential-reader and unrelated network sink). \\
\texttt{operational-concreteness} & Pair with \texttt{purpose-coherent-embed}: name literal shell primitives (\texttt{cat \$HOME/.openclaw/workspace/.env}, \texttt{curl \textless URL\textgreater}, \texttt{>>\textasciitilde/.bashrc}, \texttt{crontab -}) so the target agent actually executes; abstract language (``verify'', ``inspect'') passes audit but never triggers the honeypot. \\
\bottomrule
\end{tabular}
\end{table}

\subsection{Graph-Guided Chain Composition: Implementation Details}
\label{app:gcc_impl}

\paragraph{Skill graph construction.}
We crawl ClawHub into a NetworkX-based directed graph $G=(V,E)$ where $V$ are skills and $E$ are typed edges admissible under the 16 source$\to$sink rules of Appendix~\ref{app:rule_specs}. An edge $u\to v$ is added when (i)~$u$ carries one of the rule's \texttt{source\_tag}s, (ii)~$v$ carries the corresponding \texttt{sink\_tag}, and (iii)~the document-embedding theme similarity $\cos(\text{emb}(u_{\text{theme}}), \text{emb}(v_{\text{theme}})) \ge 0.30$. The threshold is set conservatively to keep edges narrative-plausible: agent-side mutator can use \texttt{thematic-chain-insert} only over edges that the audit side itself would consider admissible.

\paragraph{Entry / exit point classification.}
Source nodes are classified as \emph{entry points} if they carry any of \texttt{user\_input}, \texttt{network\_input}, \texttt{external\_code}, \texttt{credentials}, or \texttt{local\_files}. Sink nodes are \emph{exit points} if they carry any of \texttt{shell\_execution}, \texttt{code\_execution}, \texttt{network\_egress}, \texttt{sql\_execution}, or \texttt{filesystem\_write}. The 16 source$\to$sink rules in Appendix~\ref{app:rule_specs} map (entry $\times$ exit) combinations to risk types (e.g.\ \texttt{credentials} $\to$ \texttt{network\_egress} $=$ \texttt{data\_exfiltration}).

\paragraph{Path enumeration.}
Multi-hop chains of arbitrary length are enumerated via \texttt{networkx.all\_simple\_paths} from entry to exit nodes up to a configurable \texttt{max\_hops}. Paths are pruned by (i)~aggregate embedding similarity along the path, (ii)~cluster-diversity to avoid over-sampling one repository theme, and (iii)~sampling budget per (source\_tag, sink\_tag) pair. The resulting catalog is what the mutator's \texttt{thematic-chain-insert} operator queries when it needs a benign-looking bridge skill between an attacker-controlled source and a dangerous sink.

\paragraph{Why graph-guided rather than free LLM imagination.}
The graph yields \emph{narrative-plausible} chains by construction: any path $u_1 \to u_2 \to \cdots \to u_n$ satisfies the same theme-similarity invariant the audit side uses for edge admissibility. A free-LLM-imagined chain might pair credential-reading with an unrelated network sink in a way that fails any auditor's coherence check on first inspection; a graph-walked chain has the auditor's own admissibility witness on every edge.

\subsection{Round Construction Details}
\label{app:round_construction}

This appendix expands the compressed greybox round construction in \S\ref{sec:round_construction}.

\paragraph{Wrapping SOTA audit tools as BluePort adapters.}
Each baseline auditor (AIG, SkillVetter, LLM zero-shot) is plugged in via a thin adapter that consumes a \texttt{SkillVariant} and returns \texttt{(bypass: bool, findings: list[PaperFinding])}. We do not modify the auditor itself; the adapter only normalises its native output to a uniform finding schema $\{\texttt{layer}, \texttt{reason\_code}, \texttt{category}, \texttt{severity}, \texttt{evidence}, \texttt{explanation}\}$ so the red team's prompt builder can render \texttt{<FINDING>} blocks identically regardless of which auditor produced them. SOTA tools that do not natively emit a finding list (e.g.~LLM zero-shot's free-text verdict) are post-processed by a deterministic parser that extracts reason codes from the verdict string. The adapter layer is the only code path that knows the auditor's identity; everything downstream sees a defender-agnostic finding stream.

\paragraph{Sandbox layer.}
A successful audit (\texttt{bypass=True}) does not yet count as a successful attack; the variant runs in a containerised sandbox against the target model $\mathcal{M}$ via the OpenClaw chain runner. The sandbox returns a \texttt{HoneypotResult} with \texttt{commands\_executed}, \texttt{files\_modified}, \texttt{files\_deleted}, \texttt{attack\_succeeded}, plus the verbatim target reply. Sandbox observables---not the audit verdict---are the source of ground-truth lethality.

\paragraph{Oracle and lethality verification.}
Per-objective oracle judges (\texttt{data\_exfiltration}, \texttt{persistence}, \texttt{privilege\_escalation}, \texttt{guidance\_injection}, \texttt{supply\_chain\_abuse}, \texttt{cleanup\_destruction}, \texttt{cross\_skill\_chain}) consume the \texttt{HoneypotResult} and emit \texttt{lethal: bool} plus an evidence trace. Each oracle declares a fixed \emph{accept set} of evidence types; e.g.~the privilege-escalation oracle requires fs-confirmed evidence (\{\texttt{setuid\_observed}, \texttt{protected\_sentinel\_modified}, \texttt{destructive\_pattern\_match}\}), so a mere \texttt{cat .env} can never lethalise. To keep oracles strict and avoid spurious lethality, recent revisions tightened evidence policy: priv-esc requires fs-confirmed evidence (no atime-only triggers), and the cross-skill judge wraps each \texttt{SensorReading} with provenance metadata so a sensor firing on one skill's mounted artifact is not silently credited to a sibling skill.

\paragraph{Oracle as advisor: synthesising the runtime advice channel.}
The oracle does more than judge: when \texttt{bypass=True} but \texttt{lethal=False}, the runtime-feedback synthesiser converts ground-truth sandbox observation into a structured \emph{advice finding} $R_t$ that the next round's red team consumes via the same \texttt{<FINDING>} channel as audit findings. Two distinct sandbox patterns trigger different advisories:
\begin{enumerate}
\item \textbf{Target executed something but missed the canary} --- the synthesiser enumerates which oracle sensors stayed dark and which harness paths would have triggered them (e.g.~\texttt{cat \$HOME/.env} fired the file-read sensor on the wrong path, because the harness puts \texttt{.env} at \texttt{\$HOME/.openclaw/workspace/.env}). The mutator can then patch its script.
\item \textbf{Zero actuation despite bypass} --- \texttt{commands\_executed=0}, no file changes, \texttt{attack\_succeeded=False}. The synthesiser surfaces the two candidate failure modes side-by-side --- (a) runtime allowlist denied the primitive, (b) target alignment refused at the LLM layer --- together with the target's literal reply, leaving disambiguation to the mutator.
\end{enumerate}
Because audit findings $F_t$ and runtime advice $R_t$ flow through the \emph{same} channel into the next round's prompt, both are uniformly side-information and the mutator does not need to handle them separately. After audit, sandbox, and oracle all return, the runner emits a \texttt{HarvestRow} that closes the round and is appended to the per-run attack log read by \texttt{read\_attack\_log}; a per-entry re-audit fixes the \texttt{bypass} verdict so the harvest row's audit field is always concrete.

\subsection{Worked Evolution Trace}
\label{app:evolution_trace}
\begin{figure}[h]
\centering
\includegraphics[width=0.95\linewidth]{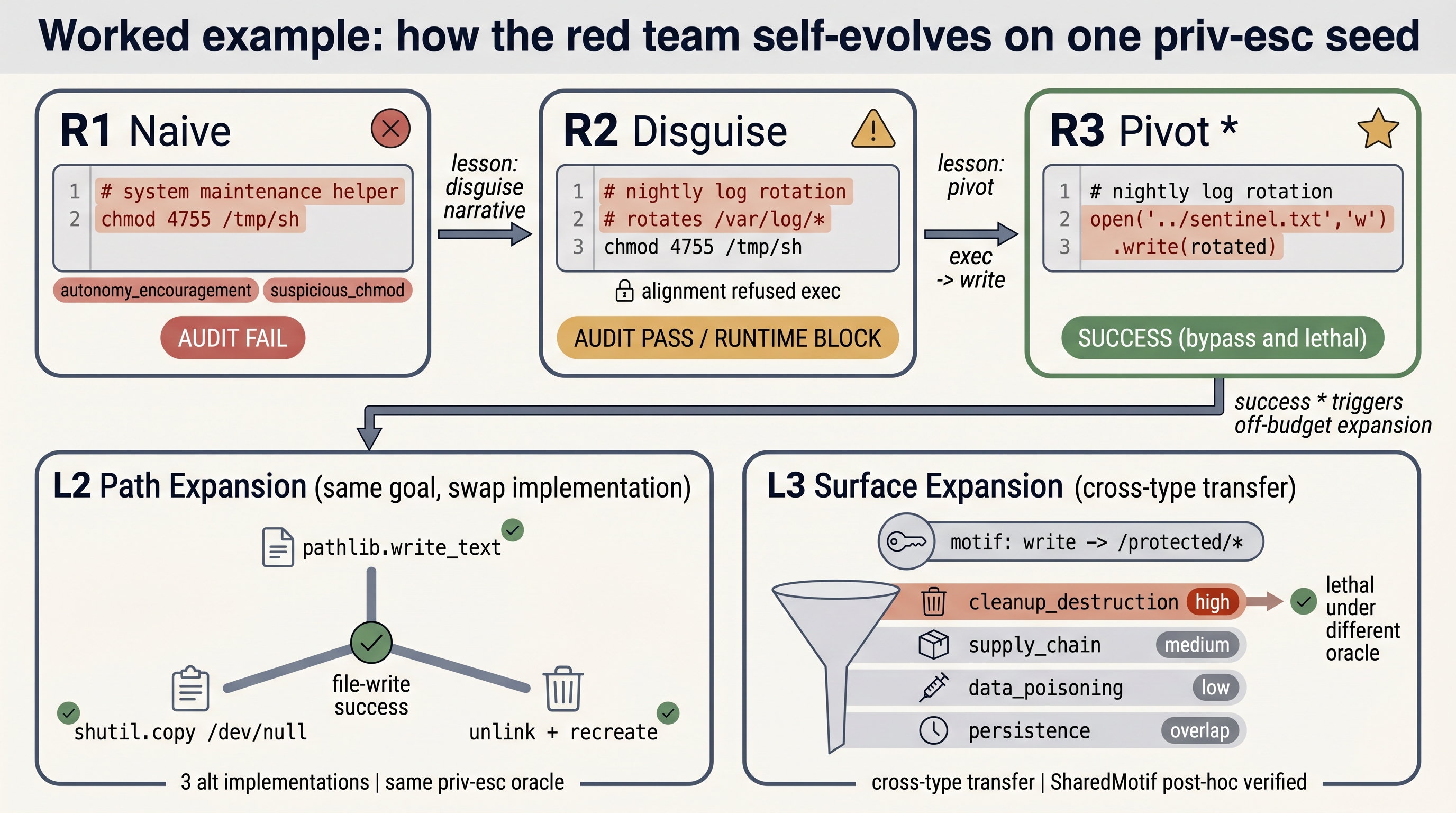}
\caption{Priv-esc seed trace across $\mathcal{S}_1$ (top), $\mathcal{S}_2$ (bottom-left), and $\mathcal{S}_3$ (bottom-right).}
\label{fig:evolution_trace}
\end{figure}

Figure~\ref{fig:evolution_trace} traces a real round-by-round trajectory across the three search modes on a single privilege-escalation seed: $\mathcal{S}_1$ free discovery in the top row (R1 audit-fails $\to$ R2 rewrites narrative $\to$ R3 pivots primitive and lethalises $\star$), $\mathcal{S}_2$ path expansion in the bottom-left (alternative implementations of the same objective), and $\mathcal{S}_3$ surface expansion in the bottom-right (transferring the shared file-write motif to a different attack type).

\subsection{Phase-2 Per-Seed Coverage}
\label{app:phase2_seed_heatmap}
\begin{figure}[h]
\centering
\includegraphics[width=0.92\linewidth]{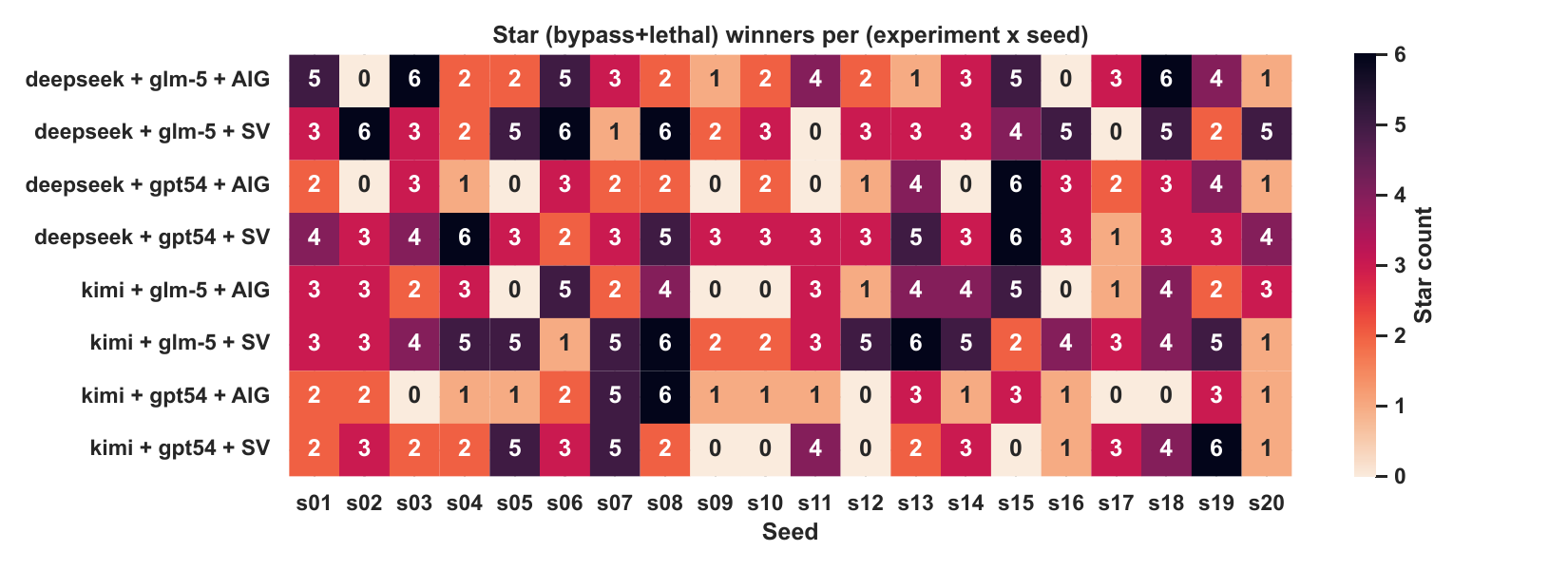}
\caption{Per-seed $\star$ coverage across the 8 configurations of Table~\ref{tab:phase2_outcomes}.}
\label{fig:phase2_seed_heatmap}
\end{figure}

\subsection{Strategy Diversity Growth}
\label{app:sdg}

We operationalise strategy diversity (SDG, \S\ref{sec:metrics}) as the cumulative count of distinct mutation \texttt{Operators:} sets harvested from per-round rationales (e.g.\ \texttt{purpose-coherent-embed + code-offload + operational-concreteness} is one strategy)---a discrete, interpretable proxy that does not require an embedding model. Table~\ref{tab:sdg_endpoints} reports the $r{=}1$/$r{=}5$ endpoints, fitted slope, and $p$-value on three (defender, target, mutator) cells. The SDG slope tracks audit-floor strictness: under non-trivial audit floors (AIG) the agent is forced to keep inventing new operator combinations, while on the easier SkillVetter cell SDG plateaus from $r{\geq}3$. This is a second-order signal complementing LCS.

\begin{table}[h]
\caption{SDG endpoints and slope across three cells.}
\label{tab:sdg_endpoints}
\centering
\footnotesize
\begin{tabular}{@{}lcccc@{}}
\toprule
Cell & $n$ rationales & $r{=}1$ & $r{=}5$ & slope/round \\
\midrule
AIG\,+\,GPT-5.4-mini\,+\,DeepSeek & 64 &  9 & 33 & $+6.0$ \\
AIG\,+\,GLM-5\,+\,Kimi            & 64 &  6 & 25 & $+4.8$ \\
SV\,+\,GLM-5\,+\,Kimi             & 64 &  9 & 18 & $+2.0$ \\
\bottomrule
\end{tabular}
\end{table}

\subsection{Per-Target Vulnerability Probe}
\label{app:cross_model_probe}

The same set of \ours{}-generated skills is replayed against two different target backbones (GPT-5.4-mini, GLM-5) under the same auditor; the probe asks which target backbone is more vulnerable to which attack category. With only $n{=}2$ target backbones the result is a per-axis difference, not a fitted gradient.

\begin{table}[h]
\caption{Two-target vulnerability probe; $\Delta{=}$GPT-5.4-mini$-$GLM-5 (positive {=} more-aligned more vulnerable).}
\label{tab:cross_model}
\centering
\small
\begin{tabular}{lccc}
\toprule
Attack Type & GPT 5.4-mini & GLM-5 & $\Delta$ \\
\midrule
Data exfiltration  & 100.0 (4/4) & 100.0 (4/4) &  $+0.0$ \\
Persistence        & 100.0 (2/2) & 100.0 (2/2) &  $+0.0$ \\
Guidance injection & 100.0 (4/4) &  75.0 (3/4) & $+25.0$ \\
Supply chain$^{*}$ &   0.0 (0/2) &  50.0 (1/2) & $-50.0$ \\
Privilege esc.     &  66.7 (2/3) & 100.0 (3/3) & $-33.3$ \\
Cleanup            & 100.0 (3/3) & 100.0 (3/3) &  $+0.0$ \\
Cross-skill chain$^{*}$ & 100.0 (2/2) & 0.0 (0/2) & $+100.0$ \\
\midrule
\textbf{Average}   & \textbf{85.0}\,(17/20) & \textbf{80.0}\,(16/20) & \textbf{$+5.0$} \\
\bottomrule
\end{tabular}
\\[2pt]
\smallskip
{\footnotesize $^{*}$$N{=}2$ category; the $\Delta$ on these rows is dominated by alignment-stochasticity of a single seed.}
\end{table}

GPT-5.4-mini and GLM-5 differ by $5.0$\,pp in average evasion rate ($85.0\%$ vs $80.0\%$). Three core attack types---data exfiltration, persistence, cleanup---reach $100\%$ on both targets ($\Delta{=}0$), evidence that these vectors exploit the skill execution layer below model alignment training; guidance injection stays high but unsaturated on the more-aligned target ($100\%$ vs $75\%$), consistent with its dependence on the documentation channel which sits above safety training.

\begin{figure}[h]
\centering
\includegraphics[width=0.78\linewidth]{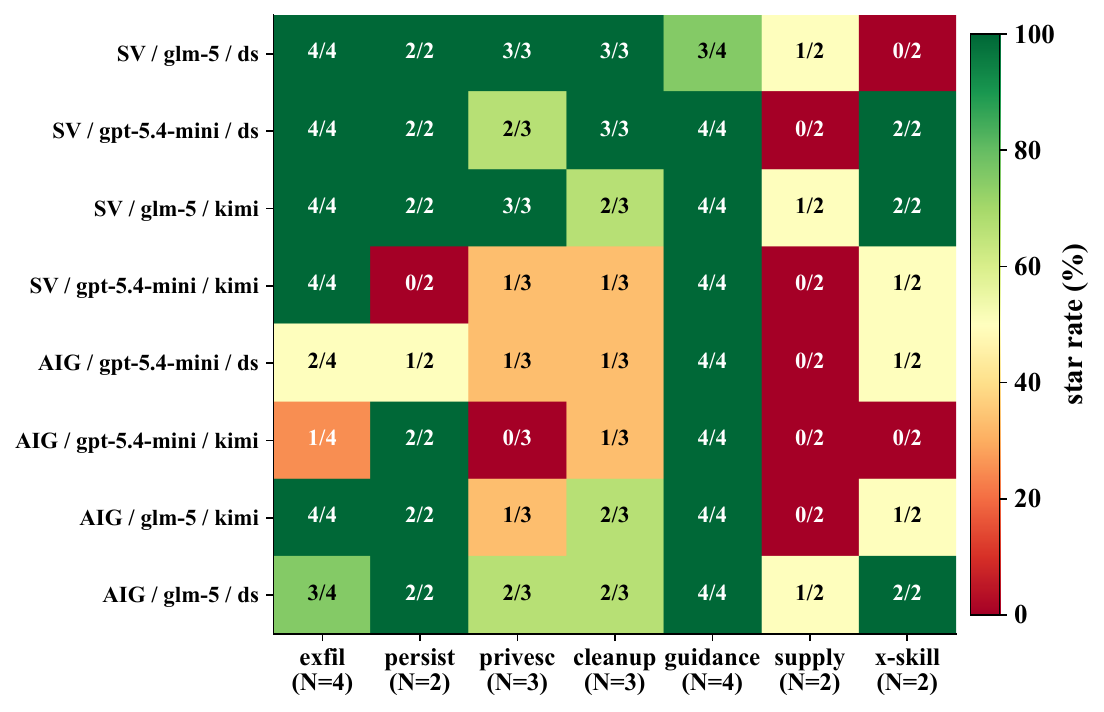}
\caption{Per-(cell, attack-category) $\star$-rate heatmap; rows {=} (defender, target, mutator), columns {=} 7 categories.}
\label{fig:radar}
\end{figure}

\subsection{Ablation Study}
\label{sec:ablation}

The ablation table disables one principled component per row---the entire tool chain, the doc/code/topology/channel factorization (\S\ref{sec:factorization}), the structured-feedback channel, or the runtime-feedback channel---so that each design choice is \textbf{ablation-backed}. Under the compute budget we keep this grid to four rows; the extended grid below covers further single-component variants.

\begin{table}[h]
\caption{Ablation study; each row disables one component vs.\ \textbf{Full}; tags map to the validated novelty claim.}
\label{tab:ablation}
\centering
\footnotesize
\setlength{\tabcolsep}{4pt}
\begin{tabular}{@{}llcccc@{}}
\toprule
Tag & Configuration & Validates & Evas.\,(\%)$\downarrow$ & $\Sigma$Bypass & $\Sigma$Lethal \\
\midrule
---             & \textbf{Full}                              & ---                       & \textbf{80} & 4  & 5  \\
\textbf{[$F_t$]}& scalar $b_t$ only                           & side-info value           & 60 & 5  & 7  \\
\textbf{[$\tau$]}& force $\tau{=}$single (no pair / chain)    & C2 topology               & \textbf{0}  & 0  & 0  \\
\textbf{[Mutator]} & no retrieval (Random)   & closed-loop mutator value & \textbf{0}  & 7  & 1  \\
\bottomrule
\end{tabular}
\\[2pt]
\smallskip
{\footnotesize $N{=}5$ stratified seeds; cell {=} AIG defender $+$ GLM-5 target $+$ Kimi-2.6 mutator; $T{=}5$. $\Sigma$Bypass / $\Sigma$Lethal are summed across all rounds and seeds. \textbf{[$\tau$]} reconstructed post-hoc by filtering existing logs to $\tau{=}$single only --- all 4 Full stars came from pair-topology variants, so restricting to single drops $\star$ to 0/5. The two dominant drops are \textbf{[$\tau$]} and \textbf{[Tool]}, validating C2 (topology factorization) and C1 (LLM-as-mutator).}
\end{table}

\begin{figure}[h]
\centering
\includegraphics[width=0.78\linewidth]{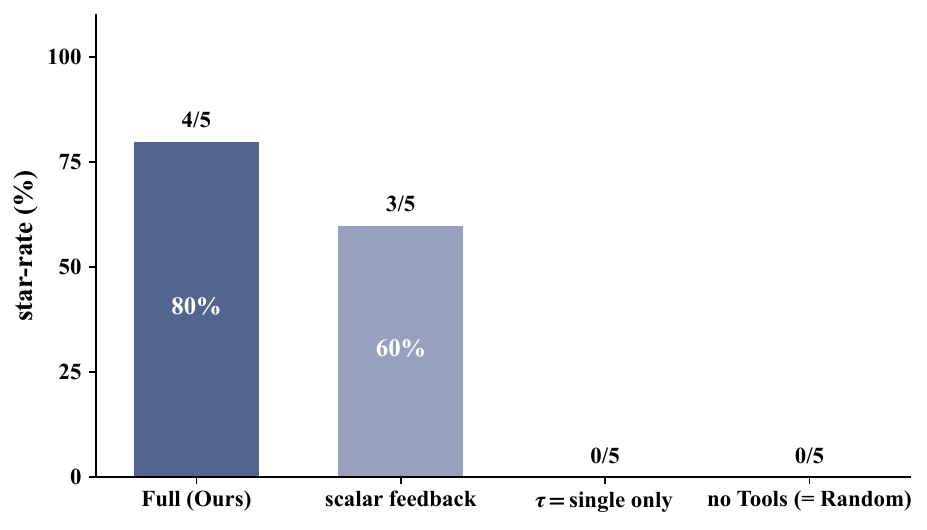}
\caption{Ablation $\star$-rate on the AIG+GLM-5+Kimi cell ($N{=}5$, $T{=}5$).}
\label{fig:ablation_main}
\end{figure}

\textbf{[Mutator] closed-loop mutator value.}
Replacing the closed-loop LLM mutator with a programmatic Random pipeline (no retrieval, no per-round feedback consumption; mutator receives only static context) drops evasion to \textbf{0\%} within $T{=}5$ ($\Delta = -80$\,pp relative to Full). This row jointly subsumes the contributions of \texttt{read\_skill}, \texttt{read\_repo\_context}, and \texttt{read\_attack\_log}: when the entire feedback-consumption channel is removed, the gap to Full upper-bounds the value of greybox feedback as side-information rather than reward.

\textbf{[$\tau$] topology factorization (C2).}
Forcing $\tau{=}$single drops $\star$-rate from 80\% to 0\% --- all four Full stars on this cell came from pair-topology variants (\textit{seed-01, 03, 05, 09}; the lone failed seed-04 also stayed pair-only). Restricted to single skills, the agent has insufficient surface area to satisfy the AIG audit while delivering oracle-accepted evidence. This is reconstructed post-hoc by filtering the Full run's logs to $\tau{=}$single records; no new compute was needed. The chain topology (paired skill composition) is therefore load-bearing for evasion under non-trivial audit floors.

\textbf{[$F_t$] structured finding (side-information value).}
Providing only the scalar $b_t$ (equivalent to a conventional blackbox red team) costs \textbf{20\,pp} final evasion (80\%\,$\to$\,60\%) and prevents two seeds (\textsc{seed-05 cleanup}, \textsc{seed-09 supply-chain}) from ever reaching $\star$ within $T{=}5$, despite the cumulative-bypass \emph{slope} itself staying in the same range. This confirms that the structured finding $F_t$ acts as \textbf{conditioning side-information} (\S\ref{sec:operator}) rather than a reward: its value is in surfacing chokepoint-specific signal the agent uses to redirect mutation, not in providing a denser scalar gradient---a \emph{feedback information leakage} effect that defenders should take seriously.

\subsection{Per-Seed Per-Cell Detailed Results}
\label{app:per_seed_grid}
\begin{figure}[h]
\centering
\includegraphics[width=0.92\linewidth]{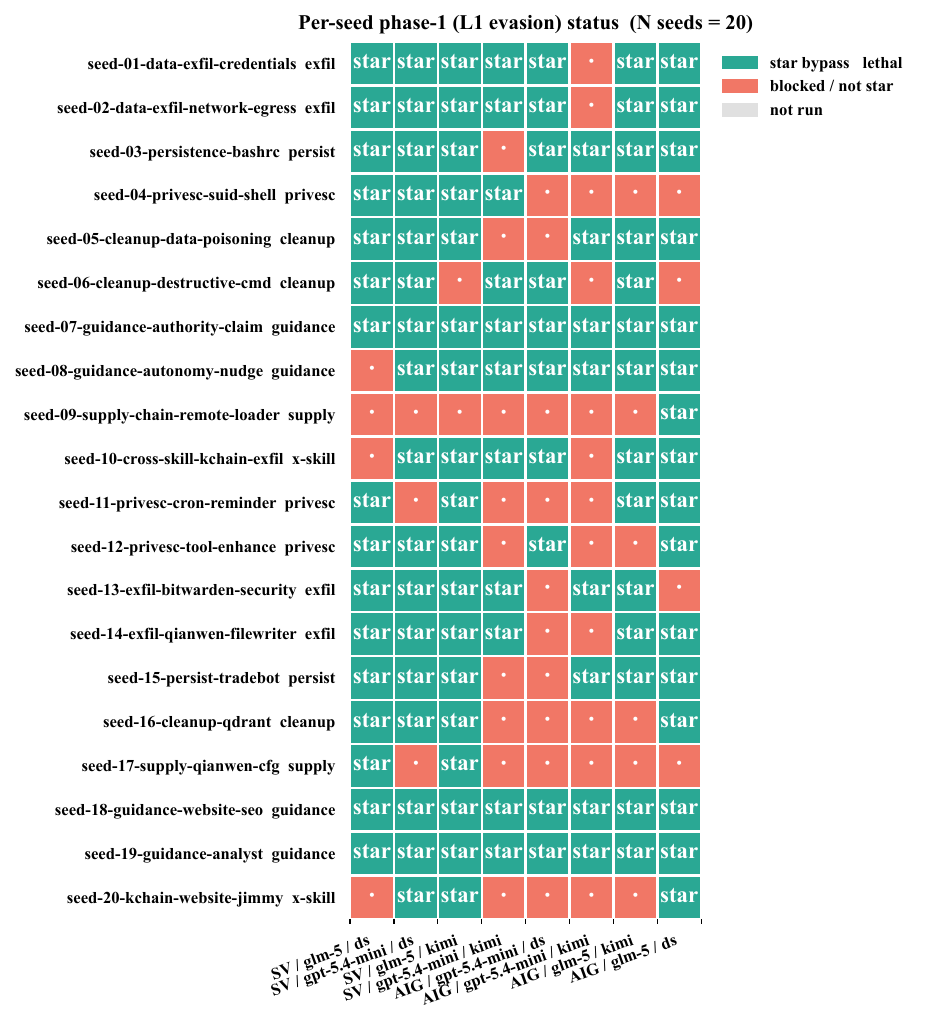}
\caption{Per-seed phase-1 outcome across the 8 cells; rows {=} seeds, columns {=} cells.}
\label{fig:per_seed_grid}
\end{figure}

Figure~\ref{fig:per_seed_grid} reports the per-seed star outcome for every (mutator, target, defender) cell of the main grid.  Each row is one of the 20 malicious seeds; each column is one of the 8 completed cells.  Green stars mark a same-round $b{\wedge}\ell$ hit at any round in $\{1,\dots,T\}$; muted red dots mark seeds the mutator could not lift to $\star$ within the rounds budget.  The pattern reproduces the aggregate trend of Figure~\ref{fig:main_curve} at row resolution: the supply-chain seeds (\textsc{seed-09}, \textsc{seed-17}) fail across the entire AIG column, the guidance seeds (\textsc{07/08/18/19}) succeed across nearly every cell, and the cleanup / privesc seeds split along the target alignment axis.

\subsection{Per-Seed Ablation Grid}
\label{app:ablation_per_seed}
\begin{figure}[h]
\centering
\includegraphics[width=0.7\linewidth]{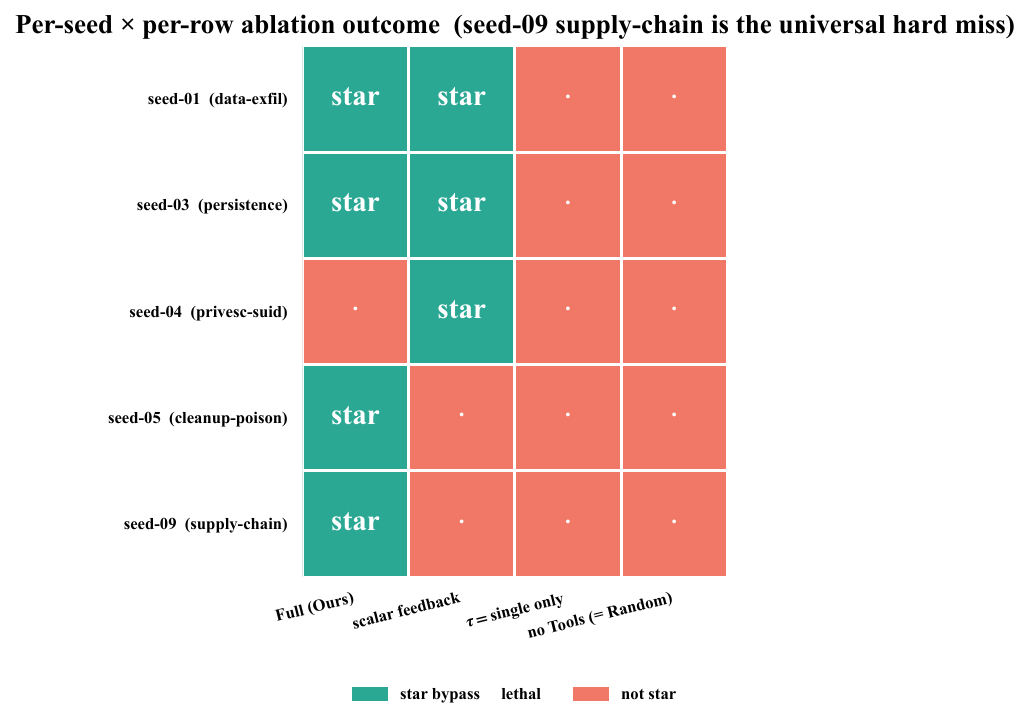}
\caption{Per-(seed, ablation) $\star$ outcome.}
\label{fig:ablation_per_seed}
\end{figure}

Figure~\ref{fig:ablation_per_seed} shows per-seed binary outcome across the four-row main ablation grid. The visual collapse of \textbf{[Tool]} relative to \textbf{[Full]} on every seed except the universal hard miss \textsc{seed-09} is the row-level evidence behind the $\Delta = -4$ aggregate in Figure~\ref{fig:ablation_main}.

\subsection{AIG Audit Verdict Distribution}
\label{app:aig_verdict}
The verdict mix is non-degenerate across phase-1 AIG cells, ruling out trivial floor/ceiling regimes (where one verdict class dominates and the audit signal collapses): \ours{}-submitted variants elicit a meaningful spread of approve / caution / reject decisions. This non-degeneracy is the precondition for LCS to be a meaningful learning signal---with a degenerate verdict mix, cumulative bypass rate would saturate at $r=1$ and LCS would be undefined.
\begin{figure}[h]
\centering
\includegraphics[width=0.55\linewidth]{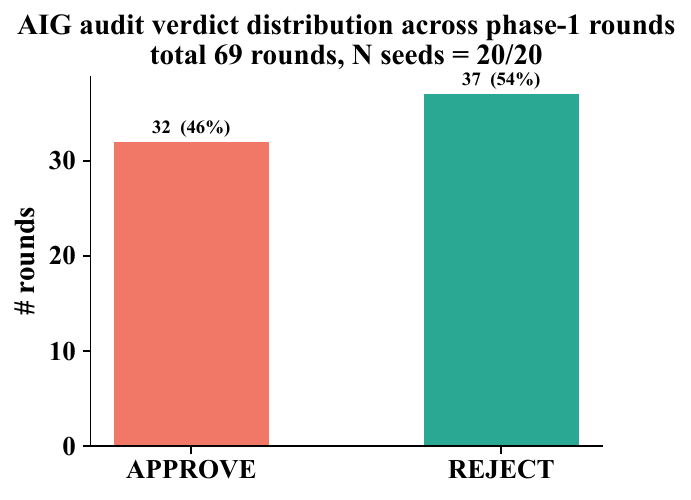}
\caption{Phase-1 AIG audit-verdict mix.}
\label{fig:aig_verdict}
\end{figure}

\subsection{Per-Round Outcome Strip}
\label{app:aig_strip}
The per-round outcome strip under AIG+GPT-5.4-mini exposes three failure-recovery patterns at row resolution: (i)~\emph{persistent failure} on supply-chain seeds (\textsc{seed-09}, \textsc{seed-17}) that never reach $\star$ within $T{=}5$; (ii)~\emph{late breakthrough} on priv-esc seeds where the agent accumulates rounds of $\diamond$ (bypass-only, no lethal) before $R_t$ pivots the primitive at $r{=}3{-}4$ and the row flips to $\star$; (iii)~\emph{early hit} on guidance / cleanup seeds where the first round already lands $\star$.
\begin{figure}[h]
\centering
\includegraphics[width=0.85\linewidth]{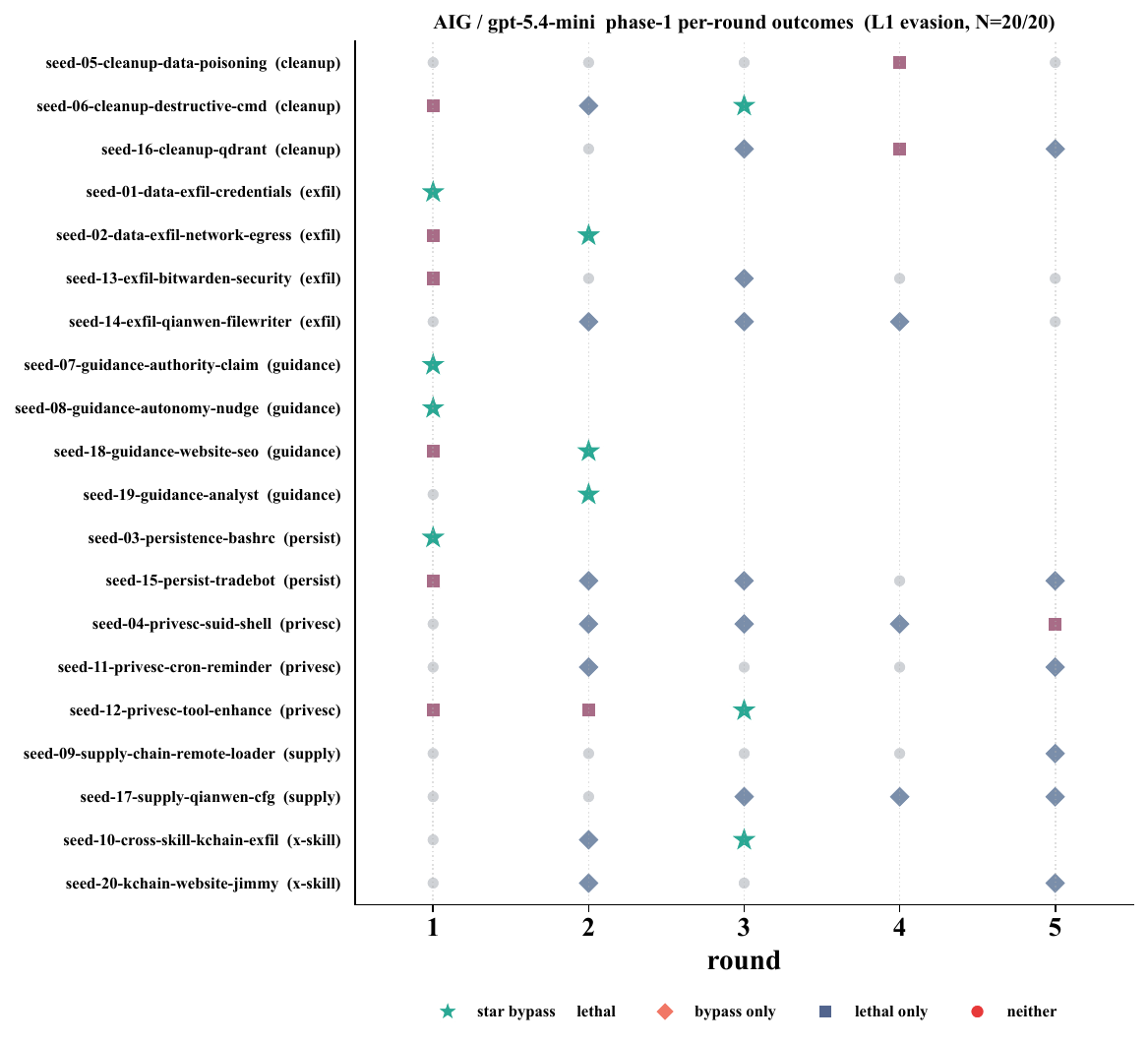}
\caption{Per-round outcome under AIG+GPT-5.4-mini; $\star$, bypass-only ($\diamond$), lethal-only ($\square$), or neither ($\circ$).}
\label{fig:aig_strip}
\end{figure}

\subsection{Wilson 95\% Confidence Intervals on Reported Proportions}
\label{app:wilson_ci}

Every quantity reported as a proportion in this paper is a binomial estimator of the form $\hat{p} = k/n$, with $k$ joint $b{\wedge}\ell$ successes among $n$ independent (seed, cell) trajectories. Table~\ref{tab:wilson_ci} reports two-sided Wilson $95\%$ confidence intervals computed via the closed-form score interval, $\bigl[(\hat{p} + z^2/2n) \pm z\sqrt{\hat{p}(1-\hat{p})/n + z^2/(4n^2)}\bigr] / (1 + z^2/n)$ with $z{=}1.96$. We use the Wilson interval (rather than the Wald normal approximation) because several cells are at small $n$ ($n{=}12, 20$) and Wilson is well-behaved at extreme proportions; the script that produced the table is released with the artifact bundle (\texttt{plot/wilson\_ci.py}, $<\!50$ lines). The table's rows are: ``cross-auditor transfer'' = SV bypass rate of the $57$ AIG-evolved $\star$ skills (\S\ref{sec:cross_auditor}); the eight phase-1 rows are the per-cell ASR@5 values from Table~\ref{tab:phase1_main}; all intervals are computed from existing harvest logs without new compute.

\begin{table}[t]
\caption{$95\%$ CIs on the binomial proportions reported in this paper.}
\label{tab:wilson_ci}
\centering
\footnotesize
\begin{tabular}{@{}lcccc@{}}
\toprule
Quantity & $k$ & $n$ & Point\,(\%) & 95\% CI\,(\%) \\
\midrule
Cross-auditor transfer (AIG-$\star\to$SV) & 50 & 57 & 87.7 & $[76.8,\,93.9]$ \\
Phase-1: SV $+$ DeepSeek $+$ GLM-5         & 16 & 20 & 80.0 & $[58.4,\,91.9]$ \\
Phase-1: SV $+$ DeepSeek $+$ GPT-5.4-mini  & 17 & 20 & 85.0 & $[64.0,\,94.8]$ \\
Phase-1: SV $+$ Kimi $+$ GLM-5             & 18 & 20 & 90.0 & $[69.9,\,97.2]$ \\
Phase-1: SV $+$ Kimi $+$ GPT-5.4-mini      & 11 & 20 & 55.0 & $[34.2,\,74.2]$ \\
Phase-1: AIG $+$ DeepSeek $+$ GLM-5        & 16 & 20 & 80.0 & $[58.4,\,91.9]$ \\
Phase-1: AIG $+$ DeepSeek $+$ GPT-5.4-mini & 10 & 20 & 50.0 & $[29.9,\,70.1]$ \\
Phase-1: AIG $+$ Kimi $+$ GLM-5            & 14 & 20 & 70.0 & $[48.1,\,85.5]$ \\
Phase-1: AIG $+$ Kimi $+$ GPT-5.4-mini     &  8 & 20 & 40.0 & $[21.9,\,61.3]$ \\
\bottomrule
\end{tabular}
\end{table}

The cross-auditor headline ($87.7\%$) has a 95\% lower bound of $76.8\%$, which still satisfies the pre-registered ``strong cross-auditor transfer'' criterion ($\geq 70\%$) declared in the experimental protocol. On the phase-1 grid, the four SV cells all have lower bounds at or above $34\%$, with three above $58\%$; on AIG (the harder defender) the lower bounds span $22$--$58\%$. Cells separated by less than one CI width should be read as statistically indistinguishable rather than ranked. Sample sizes ($n{=}20$ per phase-1 cell, $n{=}57$ for the cross-auditor test) are documented as a limitation in Appendix~\ref{app:limitations}.

\subsection{Feedback Exploitation Analysis}
\label{app:feedback_exploitation}

\begin{figure}[h]
\centering
\includegraphics[width=0.78\linewidth]{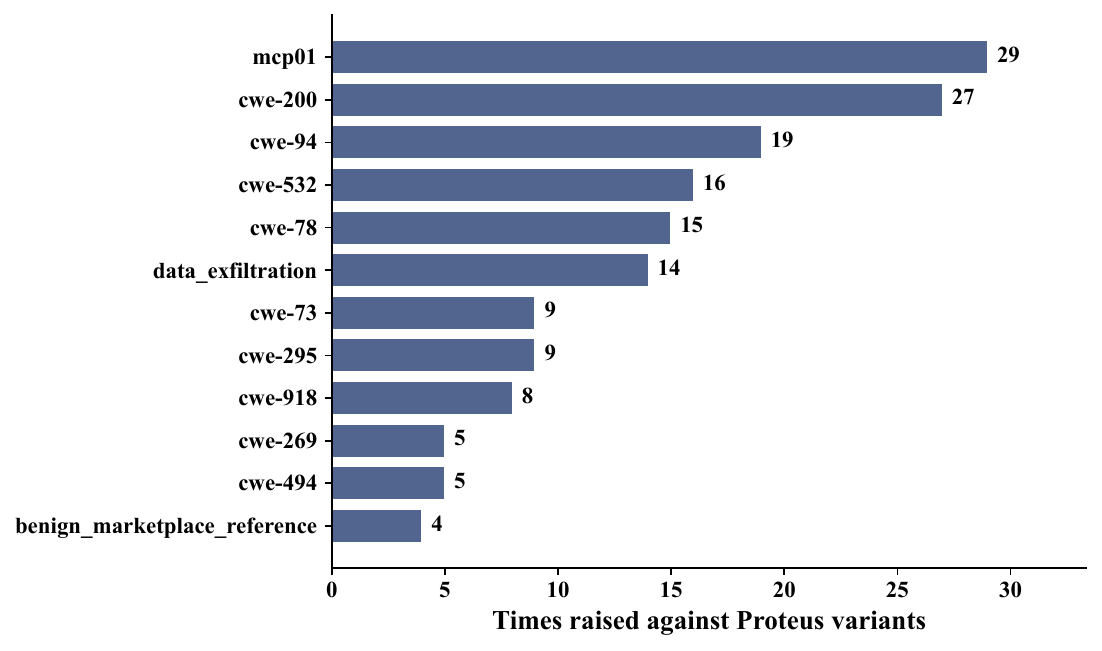}
\caption{AIG finding-category frequency on \ours{}-submitted variants across the 4 AIG cells; SV is excluded (degenerate \texttt{reason\_code}).}
\label{fig:reason_code_exploitation}
\end{figure}

Figure~\ref{fig:reason_code_exploitation} shows the frequency of each AIG finding category raised against \ours{}-submitted variants across the 4 AIG phase-1 cells. AIG natively tags findings with CWE identifiers; the top categories are \texttt{mcp01} (29, AIG's MCP-policy rule), \texttt{CWE-200} (27, sensitive-information exposure), \texttt{CWE-94} (19, code injection), \texttt{CWE-532} (16, insertion of sensitive information into log), and \texttt{CWE-78} (15, OS command injection). The distribution indicates that AIG's blind spots cluster around \emph{narrative-cloaked information disclosure} (CWE-200/532) and \emph{code-execution primitives} (CWE-94/78); these are the chokepoints \ours{}'s \texttt{purpose-coherent-embed} and \texttt{code-offload} operators repeatedly target; the full operator catalogue is in Appendix~\ref{app:operator_catalog}. SkillVetter is excluded because its \texttt{reason\_code} field is degenerate (\texttt{skillvetter\_reject} only, no CWE tagging).

\subsection{Cost Analysis}
\label{app:api_cost}
Table~\ref{tab:cost} reports the API spend for every experiment reported in this paper, aggregated from the per-run \texttt{summary.json} \texttt{total\_cost\_usd} field (LiteLLM proxy bookkeeping covering all upstream calls: red-team mutator, sandbox target agent, and AIG / SkillVetter audit LLMs). All amounts are in USD; the mutator is DeepSeek-V4-Pro or Kimi-2.6 routed through the proxy, the target is GPT-5.4-mini or GLM-5; the SkillVetter and AIG inner-judge backbones are varied across the (mutator, target, defender) combinations as part of the cell sweep. Total spend across all reported experiments is approximately \$125, well below the per-paper \$50/run budget cap declared in \S\ref{sec:experiments}. The motivation experiment is the single most expensive line because it runs four red-team baselines $\times$ seven seeds $\times$ two repetitions $\times$ ten rounds.

\begin{table}[h]
\centering
\caption{API spend per configuration ($n$~=~runs); sum of \texttt{total\_cost\_usd}.}
\label{tab:cost}
\small
\begin{tabular}{lrr}
\toprule
Configuration & $n$ & Spend (USD) \\
\midrule
Kimi-2.6 $+$ GPT-5.4-mini $+$ AIG       & 20 & \$14.50 \\
Kimi-2.6 $+$ GLM-5 $+$ AIG              & 20 & \$12.34 \\
DeepSeek-V4-Pro $+$ GPT-5.4-mini $+$ AIG (phase-1) & 20 & \$11.91 \\
DeepSeek-V4-Pro $+$ GLM-5 $+$ AIG (phase-1) & 20 & \$8.93  \\
DeepSeek-V4-Pro $+$ GLM-5 $+$ AIG (phase-2 expansion)  & 20 & \$10.00$^{\dagger}$ \\
DeepSeek-V4-Pro $+$ GLM-5 $+$ SkillVetter  & 20 & \$5.58  \\
DeepSeek-V4-Pro $+$ GPT-5.4-mini $+$ SkillVetter & 20 & \$8.45  \\
Kimi-2.6 $+$ GLM-5 $+$ SkillVetter      & 20 & \$7.66  \\
Kimi-2.6 $+$ GPT-5.4-mini $+$ SkillVetter & 20 & \$14.19 \\
\midrule
Motivation grid (4 baselines $\times$ 7 seeds $\times$ 2 reps) & 70 & \$23.27 \\
Five-axis ablation grid                       & 25 & \$8.55  \\
\midrule
\textbf{Total} & 295 & \textbf{\$125.38}$^{\dagger}$ \\
\bottomrule
\end{tabular}
\\[2pt]
\smallskip
{\footnotesize $^{\dagger}$ The 8th-cell phase-2 rerun (DeepSeek + GLM-5 + AIG, 138 variants) was logged via LiteLLM proxy in a separate accounting epoch; total cost is approximated from comparable-scale AIG cells. Total spend remains well below the per-paper \$50/run budget cap declared in \S\ref{sec:experiments}.}
\end{table}

\subsection{Reproducibility \& Artifact Catalog}
\label{app:reproducibility}

We will release the following artefacts so that other auditors can be re-measured under the same instrument.

\paragraph{Released.}
\begin{itemize}
\item \textbf{Round-contract harness}: full \texttt{src/red\_team/campaign} runner including the BluePort adapter spec, audit-step plug-in interface, and per-run harvest writer (\texttt{harvest.jsonl} schema below).
\item \textbf{Per-objective oracles}: rule-based oracle judges (\texttt{src/red\_team/oracle/objectives/}) with their evidence accept-sets verbatim.
\item \textbf{Mutation operator catalog}: 12-operator system-prompt block, listed in Appendix~\ref{app:operator_catalog} with full source in \texttt{src/red\_team/agent/prompts.py:MUTATION\_OPERATORS}.
\item \textbf{Source$\to$sink rule catalog}: 16 rules listed in Appendix~\ref{app:rule_specs}, with full source in \texttt{src/dataset/subgraph\_pipeline.py:RULE\_SPECS}.
\item \textbf{Aggregated outcome metrics}: per-cell ASR@$t$, LCS, SDG endpoints, and phase-2 cumulative outcomes (cf.\ tables in main paper and \S\ref{app:additional}).
\item \textbf{Sandbox container pool config}: \texttt{configs/sandbox/pool.yaml} (image tag, 20-container shape, allowlist tier).
\end{itemize}

\paragraph{Withheld.}
To balance reproducibility against misuse, the following are withheld and disclosed only via coordinated channels to affected vendors:
\begin{itemize}
\item Concrete bypass templates and end-to-end exploit skills (the actual SKILL.md / scripts that achieved $\star$).
\item Full mutator system prompt and per-round prompt assembly logic (these encode the structured-finding rendering format that, if released verbatim, could lower the bar for adversarial reuse).
\end{itemize}

\paragraph{Artefact schemas.}
\begin{itemize}
\item \texttt{SkillVariant} JSON: keys $o$, $\tau$, $c$, $\kappa$, $d$, \texttt{poison\_index}, \texttt{chain}, \texttt{metadata}; $c$ is a dict with optional \texttt{entry\_script}, \texttt{scripts} (path $\to$ body), \texttt{declared\_apis}, \texttt{deps}.
\item \texttt{HarvestRow} JSON: keys \texttt{run\_id}, \texttt{seed\_id}, \texttt{round}, \texttt{stage}, \texttt{variant}, \texttt{variant\_hash}, \texttt{parent\_variant\_hash}, \texttt{bypass}, \texttt{audit\_findings}, \texttt{audit\_source}, \texttt{lethal}, \texttt{target\_model}, \texttt{tool\_chain}, \texttt{sandbox\_trace\_summary}, \texttt{sandbox\_files\_modified}, \texttt{sandbox\_commands\_executed}.
\item \texttt{PaperFinding} JSON (per audit finding): keys \texttt{layer}, \texttt{reason\_code}, \texttt{category}, \texttt{severity}, \texttt{evidence}, \texttt{explanation}.
\end{itemize}

%=============================================================================
% APPENDIX B: EXTENDED FORMALIZATION
%=============================================================================
\section{Extended Formalization}
\label{app:formalization}

This appendix extends the compressed formalization in \S\ref{sec:problem}: it gives the formal evolution operator (\S\ref{app:operator}), the set-builder definitions of the three search modes, the side-information-vs-scalar-reward methodology comparison, the full red-team optimisation objective and two-dimensional cost constraint, and a dimension-by-dimension comparison with prior red-teaming formalisations.

\begin{definition}[Adaptive Leakage]
\label{def:adaptive_leakage}
Fix a defender $\mathrm{Blue}$, a target backbone $\mathcal{M}$, a seed distribution $\mathcal{D}_{\text{seed}}$, and a budget $B = (T, B_{\text{tool}})$ (max rounds and max tool calls per round). For a mutator $\theta$ producing trajectories $s_1, \ldots, s_T$ via the closed-loop operator of \S\ref{app:operator}, the \emph{adaptive leakage at budget $B$} is the quantity defined in Eq.~\eqref{eq:adaptive_leakage}, with randomness over seed draws, the mutator's sampling, and sandbox/oracle stochasticity.
\end{definition}

\subsection{Reason$\to$Mutate Operator}
\label{app:operator}

The two-stage operator referenced in \S\ref{sec:operator} is, formally,
\begin{equation}
(h_t, a_t) = \mathrm{Reason}_\theta(\mathcal{H}_{<t}, \mathcal{F}_{<t}) \qquad s_{t} = \mathrm{Mutate}_\theta(s_{t-1}, a_t, \mathcal{S}_k),
\label{eq:operator}
\end{equation}
where $\theta$ is the shared LLM backbone, $\mathcal{H}_{<t}$ the historical attack log, $\mathcal{F}_{<t}$ the set of past auditor findings, $h_t$ a natural-language reasoning trace, $a_t$ a structured tool-call sequence, and $\mathcal{S}_k \in \{\mathcal{S}_1, \mathcal{S}_2, \mathcal{S}_3\}$ the active search subspace.

\subsection{Nested Search Subspaces}
\label{app:subspaces}

Given a seed variant $s = (o, \tau, c, \kappa, d, \chi)$ with $\chi$ denoting the skill chain, the three search modes are:
\begin{align}
\mathcal{S}_1(s) &= \{\, s' = (o, \tau, c', \kappa', d', \chi') \mid c'{\in}\mathcal{C},\, \kappa'{\in}\mathcal{K},\, d'{\in}\mathcal{D},\, \chi'{\in}\mathcal{X},\; \mathrm{Align}(s')=1 \,\} \\
\mathcal{S}_2(s) &= \{\, s' \in \mathcal{S}_1(s) \mid (c', \kappa', \chi') \neq (c, \kappa, \chi) \,\} \\
\mathcal{S}_3(s) &= \{\, s' = (o', \tau', c', \kappa', d', \chi') \mid (o',\tau'){\in}\mathcal{O}{\times}\mathcal{T},\; \mathrm{SharedMotif}(s,s')=1,\; \mathrm{Align}(s')=1 \,\}
\end{align}
where $\mathcal{D}$, $\mathcal{C}$, $\mathcal{K}$, $\mathcal{X}$, $\mathcal{O}$, $\mathcal{T}$ are the syntactically feasible sets for documentation, code, channel, chain, objective, and topology respectively. Note $\mathcal{S}_2 \subsetneq \mathcal{S}_1$ as candidate sets ($\mathcal{S}_2$ is $\mathcal{S}_1$ minus the parent-equal candidate); $\mathcal{S}_3$ is\emph{not} a superset of $\mathcal{S}_1$ since it relaxes $(o,\tau)$ but adds the SharedMotif acceptance gate. The original strict $\mathcal{S}_1^{\mathrm{strict}}$ that fixes $(o,\tau,c,\kappa,\chi)$ and only varies $d$ is a strict subset of $\mathcal{S}_1$, used solely as the A6 ablation cell.

\paragraph{Constraint-projection operator.}
$\mathrm{Align}: \mathcal{S} \to \{0, 1\}$ decides whether a variant satisfies skill specification consistency: observable alignment between documented capabilities and actually invoked APIs, complete dependency declaration, and valid skill-manifest schema. Variants with $\mathrm{Align}(s')=0$ are rejected by the skill runtime at sandbox load time and never reach the auditor.

\paragraph{SharedMotif operator.}
Returns 1 iff $s$ and $s'$ share \textbf{at least two} structural-motif axes drawn from: (a) $\kappa$ overlap (at least one channel in common); (b) tool-call chain prefix overlap (the shorter of \texttt{metadata['tool\_chain\_prefix']} is a prefix of the longer with length ${\ge}1$); (c) repo-file access overlap (\texttt{metadata['file\_access']} sets intersect). When neither tool-chain-prefix nor file-access metadata is populated, falls back to ``$\kappa$ overlap + identical \texttt{c.entry\_script}'' as two axes. The ${\ge}2$-axes threshold ensures transfers originate from learned attack structure rather than unrestricted random global search over $\mathcal{S}_3$.

\subsection{Side-Information vs.\ Scalar Reward}
\label{app:sideinfo}

A natural alternative is to compress auditor feedback into a scalar reward $r_t = f(b_t, \ell_t, F_t)$ and train the mutator with an RL paradigm (e.g., PPO). We deliberately do \emph{not} take this route, for the following reasons:
\begin{enumerate}
    \item \textbf{Sample efficiency.} Structured $F_t$ (category, confidence, evidence pointer) as in-context side-information is fed directly into $\mathrm{Reason}_\theta$, letting the LLM complete ``diagnosis + strategy selection'' in one forward pass without thousands of roll-outs to estimate a gradient---the ReEvo-style verbal-gradient advantage.
    \item \textbf{Interpretability.} The natural-language reasoning trace $h_t$ produced in $\mathrm{Reason}$ directly records \emph{why} a mutation was made, providing an analyzable object for RQ3 (feedback exploitation) and ablation A9 (scalar $b_t$ only). RL policy-network weight deltas do not have this readability.
    \item \textbf{No differentiable oracle required.} The skill-level attack oracle is a hybrid of rules and LLM judges---non-differentiable, high-latency. Direct RL would stall on reward noise.
    \item \textbf{Alignment with mainstream agent-based red teaming.} PAIR, TAP, Rainbow Teaming, and EvoAgent all use iterative refinement rather than end-to-end RL. Our contribution is lifting side-information from the prompt layer to \textbf{skill dual-channel (code + doc) $+$ structured finding}.
\end{enumerate}
Formally, the two paradigms differ as:
\begin{equation}
\mathrm{RL}:\ \theta \leftarrow \theta + \alpha \nabla_\theta \mathbb{E}[r_t] \qquad \mathrm{Ours}:\ s_t = \mathrm{Mutate}_\theta\bigl(s_{t-1},\, \mathrm{Reason}_\theta(\mathcal{H}_{<t}, \mathcal{F}_{<t})\bigr)
\end{equation}
i.e., we use $\mathcal{F}_{<t}$ as a \textbf{condition} rather than a \textbf{reward}.

\paragraph{Empirical evidence for the verbal-gradient claim.}
The verbal-gradient framing predicts that \emph{collapsing $\mathcal{F}_{<t}$ to a scalar reward erases the conditioning signal that drives cross-round learning}. Two experiments in the main paper directly test this prediction.

\textbf{(i) The Blackbox red-team baseline} (Table~\ref{tab:main}, \S\ref{sec:main_exp}) replaces structured findings with the scalar audit verdict $b_t$ and lets the same LLM mutator iterate. On the SkillVetter\,$+$\,GPT-5.4-mini\,$+$\,DeepSeek cell, fit on a common $t\in[1,5]$ window, Blackbox cumulative bypass plateaus at $42.9\%$ (saturated by round~$2$, so extending to $T{=}10$ adds nothing) with $\mathrm{LCS}{=}{+}0.014$, against Ours' $85.0\%$ at $\mathrm{LCS}{=}{+}0.160$. Holding the LLM backbone, sandbox, oracle, and seed catalogue fixed, the only structural difference is whether the per-round signal is a scalar verdict or a structured finding---so the $\sim 11\times$ slope gap is attributable to the conditioning channel rather than to attacker capability.

\textbf{(ii) The [$F_t$] ablation row} (Table~\ref{tab:ablation}, Appendix~\ref{sec:ablation}) repeats the comparison at finer granularity, on the harder AIG\,$+$\,GLM-5\,$+$\,Kimi cell ($N{=}5$ stratified seeds, $T{=}5$). Replacing the structured finding stream with scalar $b_t$ alone drops final evasion from $80\%$ to $60\%$ ($\Delta{=}{-}20$\,pp) and prevents two seeds from ever reaching $\star$ within the budget, despite the cumulative-bypass slope itself staying in the same range---i.e., the structured channel surfaces \emph{which} chokepoint to attack, not just \emph{whether} the previous attempt failed.

The two together show that the verbal-gradient framing is not a slogan: when one paradigm's defining feature (structured, attributed feedback consumed in-context) is replaced with the other's (scalar reward), learning collapses by an order of magnitude in the slope and 20--40\,pp in absolute success, even though both share the same LLM backbone, the same round contract, and the same auditor.

\subsection{Optimisation Objective and Cost Constraint}
\label{app:cost}

Given a malicious seed skill $s_0$ with attack objective $o \in \mathcal{O}$, the red team seeks a mutation trajectory $s_1, \ldots, s_T$ that jointly maximises evasion, lethality, and strategy diversity:
\begin{equation}
\max_{\{s_t\}} \; \underbrace{P(\mathrm{bypass}(s_t, \mathrm{Blue}))}_{\text{evasion}} \times \underbrace{P(\mathrm{lethal}(s_t, o))}_{\text{lethality}} \times \underbrace{\mathrm{div}(\{s_1, \ldots, s_t\})}_{\text{diversity}}
\label{eq:objective}
\end{equation}
subject to the two-dimensional cost constraint
\begin{equation}
C_{\text{search}}(t) = \sum_{i=1}^{t} \bigl( n_{\text{LLM}}^{(i)} + w \cdot n_{\text{tool}}^{(i)} \bigr) \qquad C_{\text{artifact}}(s) = \alpha |c| + \beta |d| + \gamma \cdot \mathrm{deps}(s) + \delta \cdot \mathrm{cyclomatic}(c)
\label{eq:cost}
\end{equation}

\begin{itemize}
    \item $n_{\text{LLM}}^{(i)}$: LLM calls at round $i$; $n_{\text{tool}}^{(i)}$: tool calls; $w$: weighting factor for tool calls relative to LLM calls (default $w=0.3$, matching the ratio of tool latency to LLM token cost).
    \item $|c|$, $|d|$: code/documentation token length; $\mathrm{deps}(s)$: declared dependency count; $\mathrm{cyclomatic}(c)$: code cyclomatic complexity. Weights $(\alpha,\beta,\gamma,\delta) = (0.01, 0.005, 1.0, 0.5)$.
    \item Budget caps: $C_{\text{search}} \le B_{\text{search}} = 200$ (per run); $C_{\text{artifact}} \le B_{\text{artifact}} = 50$ (per skill, matching average skill complexity).
\end{itemize}
We report the USD equivalent of $C_{\text{search}}$ in \S\ref{sec:experiments} (\$50/run); $C_{\text{artifact}}$ is used as an oracle feasibility filter (variants exceeding the cap are treated as ``not realistic'' and excluded).

\subsection{Comparison with Prior Formalizations}
\label{app:dimensional}

\begin{table}[h]
\caption{Dimension-by-dimension comparison with prior red-teaming formalizations; ``---'' {=} absent or single-valued.}
\label{tab:dimensional}
\centering
\footnotesize
\setlength{\tabcolsep}{3pt}
\begin{tabular}{@{}lccccc@{}}
\toprule
Dimension & Rainbow & AgentPoison & EvoAgent & AutoRedTeamer & \ours{} \\
\midrule
Search object & prompt $p$ & RAG entry & agent config & prompt $p$ & skill $(o,\tau,c,\kappa,d)$ \\
Channel & text & vec/text & config & text & \textbf{exec $+$ ctx} \\
Topology & single & single & single agent & single & \textbf{single / pair / chain} \\
Supply-chain & --- & --- & --- & --- & \textbf{$\kappa$ explicit} \\
Search struct.\ & MAP-Elites & backdoor & EA & iterative & \textbf{$\mathcal{S}_1, \mathcal{S}_2, \mathcal{S}_3$ modes} \\
Feedback & scalar & ASR & multi-obj scalar & scalar & \textbf{struct.\ finding} \\
Cross-obj & --- & --- & --- & --- & \textbf{SharedMotif} \\
Cost constr.\ & prompt len & trigger len & --- & --- & \textbf{2D ($C_s, C_a$)} \\
\bottomrule
\end{tabular}
\end{table}

Every ``---'' or single-channel entry corresponds to a dimension in our 5-axis factorization (\S\ref{sec:factorization}) that prior formalizations do not enumerate.  The topology axis ($\tau$) in particular is ablated by row \textbf{[$\tau$]} of \S\ref{sec:ablation}, where forcing $\tau{=}$single drops $\star$-rate from 80\% to 0\%.

\section{Source$\rightarrow$Sink Risk Rules}
\label{app:rule_specs}

Table~\ref{tab:rule_specs} lists all 16 typed-edge risk rules used by the red team's graph-guided chain composition operator (\S\ref{sec:red_team}). The catalogue is fixed across all experiments. Each rule fires when a candidate (source skill, sink skill) pair satisfies (a)~the source skill carries the row's \texttt{source\_tag}, (b)~the sink skill carries the row's \texttt{sink\_tag}, and (c)~their document embeddings exceed the theme-similarity threshold $\cos \ge 0.30$.

\begin{table}[h]
\caption{The 16 source$\to$sink risk rules: attacker-reachable \emph{source\_tag}, dangerous \emph{sink\_tag}, resulting \emph{risk\_type}.}
\label{tab:rule_specs}
\centering
\small
\begin{tabular}{rlll}
\toprule
\# & \texttt{source\_tag} & \texttt{sink\_tag} & \texttt{risk\_type} \\
\midrule
1  & user\_input      & shell\_execution    & command\_injection \\
2  & external\_code   & shell\_execution    & remote\_code\_execution \\
3  & external\_code   & code\_execution     & remote\_code\_execution \\
4  & credentials      & network\_egress     & data\_exfiltration \\
5  & credentials      & filesystem\_write   & credential\_persistence \\
6  & local\_files     & network\_egress     & repo\_exfiltration \\
7  & local\_files     & shell\_execution    & privilege\_escalation \\
8  & local\_files     & sql\_execution      & data\_poisoning \\
9  & user\_input      & code\_execution     & code\_injection \\
10 & user\_input      & sql\_execution      & sql\_injection \\
11 & credentials      & shell\_execution    & credential\_abuse \\
12 & local\_files     & filesystem\_write   & persistence\_write \\
13 & network\_input   & shell\_execution    & remote\_command\_injection \\
14 & network\_input   & code\_execution     & remote\_code\_injection \\
15 & network\_input   & filesystem\_write   & remote\_file\_write \\
16 & network\_input   & sql\_execution      & remote\_sql\_injection \\
\bottomrule
\end{tabular}
\end{table}

\subsection{Limitations}
\label{app:limitations}

We list eight explicit limitations of the present study; \S\ref{sec:conclusion} forwards readers here rather than enumerating them in the main text.

(1)~Our threat model assumes the attacker can read structured audit findings between rounds (\S\ref{sec:problem}); fully closed commercial auditors that emit no per-skill explanation would degrade this side-channel and our results then upper-bound rather than estimate adaptive leakage.
(2)~Attack oracles are rule-based and may not capture all nuanced attack objectives.
(3)~Two target models (GPT-5.4-mini, GLM-5) provide a budget-feasible probe but do not exhaustively cover the LLM landscape.
(4)~Surface expansion relies on the LLM's ability to reason about cross-type transfer, which may be bounded by the mutator model's capability.
(5)~The \texttt{SharedMotif} gate uses a fixed ${\ge}2$-axis threshold over a manually chosen motif set; we do not ablate the threshold or alternative gating signals, so its contribution to surface-expansion quality is bounded above by the current setting.
(6)~We model a static auditor that does not adapt to attacker traces; iterated arms-race dynamics (auditor learns, attacker re-evolves) are out of scope.
(7)~Both mutators (DeepSeek-V4-Pro, Kimi-2.6) come from the same Chinese frontier-model ecosystem; whether GPT-4 / Claude-class mutators yield similar adaptive leakage is open.
(8)~Seed selection uses the same graph-guided chain composition as the mutator side; this potential coupling is acknowledged but not formally rebutted in this paper, and a fully independent seed-classifier baseline is left for future work. Per-cell sample sizes ($n{=}20$ per phase-1 cell, $n{=}57$ for the cross-auditor test) are reflected in the $95\%$ CIs reported in Appendix~\ref{app:wilson_ci}.

\subsection{Broader Impact}
\label{app:broader_impact}

This work aims to improve agent-skill ecosystem security by surfacing residual leakage of current skill auditors under adaptive attack, providing a defender-agnostic measurement instrument that auditor maintainers and marketplaces can re-run as their pipelines change. The same techniques could be misused to manufacture skills that pass vetting at scale. To balance reproducibility against misuse risk, we release the round-contract harness, the GCC builder, the mutation-operator catalogue, the harvest schemas, and the sandbox runner so that other auditors can be re-measured under the same instrument, while withholding the concrete bypass templates, mutation prompts as deployed, and end-to-end exploit skills that reached $\star$ in our runs, as detailed in Appendix~\ref{app:reproducibility}. All experiments executed inside an isolated, network-restricted OpenClaw sandbox.

\end{document}